\documentclass[conference]{IEEEtran}

\pdfoutput=1

\IEEEoverridecommandlockouts

\usepackage{color}
\usepackage[dvipsnames]{xcolor}
\usepackage{caption}
\usepackage{subcaption}

\usepackage{algorithm}
\usepackage{algorithmic}
\usepackage{amsmath}
\usepackage{mathtools}

\ifCLASSINFOpdf
   \usepackage[pdftex]{graphicx}
   \DeclareGraphicsExtensions{.pdf,.jpeg,.png}
\else
   \usepackage[dvips]{graphicx}
   \DeclareGraphicsExtensions{.eps}
\fi
\usepackage{amsmath}
\begin{document}

\title{A New Queue Discipline for Reducing Bufferbloat Effects in HetNet Concurrent Multipath Transfer}

\IEEEaftertitletext{\vspace{-2.0\baselineskip}}

\author{Benevid~Felix,~\IEEEmembership{Student Member,~IEEE,}
        Aldri~Santos,~\IEEEmembership{Member,~IEEE}
        and~Michele~Nogueira,~\IEEEmembership{Member,~IEEE}
\thanks{B. Silva, A. Santos and M. Nogueira are with the Department
of Informatics, Federal University of Paran\'a (UFPR), Curitiba,
Brazil e-mail: \{bfsilva, aldri, michele\}@inf.ufpr.br.}
\thanks{A. Santos and M. Nogueira are with Carnegie Mellon University, USA, on Sabbatical leave from UFPR.}

\thanks{{\bf Notice:} This work has been submitted to the IEEE for possible 
           publication. Copyright may be transferred without notice, after 
           which this version may no longer be accessible.}

}

\IEEEaftertitletext{\vspace{-1.0\baselineskip}}

\maketitle

\begin{abstract}

Heterogeneous wireless networks have evolved to reach application requirements for low latency and high throughput on Internet access. Recent studies have improved network performance employing the Multipath TCP, which aggregates flows from heterogeneous wireless interfaces in a single connection. Although existing proposals are powerful, coupled congestion control algorithms are currently limited because of the high variation in path delays, bandwidth and loss rate, typical from heterogeneous wireless networks, even more over concurrent multipath transmissions. These transmissions experience bufferbloat, i.e., high delays caused by long queues. Hence, to cope with the current limitations, this work presents CoDel-LIFO, a new active queue management (AQM) discipline to reduce the dropped packet ratio in the Multipath TCP congestion control mechanism. Differently from other approaches, CoDel-LIFO gives priority to the most recent packets, being then promising. This paper provides a detailed simulation analysis over congestion control algorithms by comparing CoDel-LIFO to CoDel and DropTail disciplines. Results indicate that CoDel-LIFO reduces queue drops, diminishing the impact on congestion control; improving substantially the goodput; and keeping RTT low.

\end{abstract}

\IEEEpeerreviewmaketitle

\section{Introduction}\label{sec:intro}

\IEEEPARstart{L}{ow} latency and high throughput are requirements of applications running over heterogeneous wireless networks (HetNets)~\cite{hossain:15}. In order to meet these requirements, researchers investigate new forms of network access, introducing small cells, device-to-device (D2D) communication and others~\cite{bangerter:14}. Nowadays, smartphones, tablets and notebooks are embedded with multiple radio interfaces (multihomed devices) such as WiFi, 3G/4G and Bluetooth access technologies. However, despite equipped with these multiple interfaces, multihomed devices still face limitations in keeping connection during a vertical handover (e.g. 3G to WiFI) when employing TCP. 


In order to benefit from the potential of multiple interfaces to improve communication performance, IETF leads a process to the standardization of the Multipath Transmission Protocol (Multipath TCP)~\cite{RFC6824}. Multipath TCP intends to support multiple path transmissions~\cite{lee:15,mehani:15} and its default congestion control (CC) algorithm runs in end-hosts. CC duplicates acknowledgments as signals to identify a packet loss or a network congestion. This is compatible with the majority of middle-boxes (NATs, firewalls, proxies, etc.) and it is employed in heterogeneous wireless networks to comply with different goals, such as latency reduction, throughput improvement, seamless handover and resiliency~\cite{ferlin14-2}.


Ferlin~{\em et al.}~\cite{ferlin14-2} and Arzani~{\em et al.}~\cite{arzani:14} show that Multipath TCP over heterogeneous wireless networks results in many issues. For instance, these issues are due to the different characteristics in the paths (QoS constraints), which compromises the Multipath TCP performance. The Multipath TCP degradation increases when the paths have wide delay difference caused by the bufferbloat phenomenon~\cite{ferlin14}. This phenomenon in general occurs due to large queues that absorb a huge amount of traffic in a congested path and produces high latency despite of low packet loss. The bufferbloat affects the performance of multipath transfers because the packets arrive out of order and with high delay between arrivals at the receiver buffer. 

{\bf Problem.} Recent studies show the drawbacks of bufferbloat in multipath transmission over heterogeneous wireless networks~\cite{alfredsson13}. Few approaches have attempted to mitigate these effects by adjusting Multipath TCP in end-hosts or by Active Queue Management (AQM) discipline in network links (e.g. routers, gateways or access points)~\cite{gong:14}. 
Multipath TCP adjustments include solutions as limiting the amount of data sent by congested path, dynamic adjustment of the receive window, disabling TCP slow start mechanism, and others. However, to the best of our knowledge, the AQM Controlled Delay (CoDel) and its variations, such as Flow Queue Controlled Delay (FQ-CoDel), offer in general the best results against the bufferbloat phenomenon with a significant latency reduction.

CoDel prevents bufferbloat controlling the queue filler in network routers by dropping proactively packets that has long time in queue. CoDel mechanism reduces the transmission latency, however, packet drops affect the loss-based congestion control algorithm, compromising the transmission performance. Nichols and Jacobson~\cite{nichols:12} point out the tradeoff between latency and goodput over the TCP protocol. 

{\bf Contribution.}
In face to this scenario, the contribution of this work is twofold:
$(i)$ the assessment of the {\em tradeoff  between latency and goodput over multipath transmissions} and $(ii)$ {\em CoDel-LIFO, new AQM discipline to reduce the dropped packet ratio in Multipath TCP}. This work presents an evaluation of the impact caused by CoDel in Multipath TCP congestion control, comparing it to default router queue management discipline (DropTail). Results show that CoDel has an expressive impact on congestion control algorithms. It results in a high dropped packet ratio, and this has an impact on the CC, preventing it to use full link capacity. Simulation results point out that CoDel compromises the goodput to reduce latency, while DropTail has better transmission rate, but over a cost of high RTT.

Therefore, this work also introduces {\bf CoDel-LIFO}, a AQM CoDel-based discipline to control the queue size and to reduce the impact of queue losses in multipath transmissions. CoDel-LIFO employs LIFO (Last In, First Out) as queue discipline in order to forward the packets with less sojourn time. It defines a parameter $\theta$ to control queue losses. Simulation results comparing CoDel-LIFO and CoDel with different Multipath TCP congestion controls show that CoDel-LIFO improves the goodput without compromising RTT.

{\bf Paper outline.} This paper proceeds as follows. Section~\ref{sec:mptcp} overviews the Multipath TCP architecture and presents its performance in HetNets. Section~\ref{sec:codel} describes CoDel and how it control path delay. The CoDel-LIFO is introduced in Section~\ref{sec:lifocodel}. Section~\ref{sec:evaluation} details the evaluation scenarios, metrics and shown simulation results. Section~\ref{sec:related} describes the related works. Section~\ref{sec:conclusion} concludes the paper.

\section{Multipath TCP}\label{sec:mptcp}

\begin{figure}[htbp]
    \centering
    \begin{subfigure}[b]{0.24\textwidth}
        \centering
        \includegraphics[width=\textwidth]{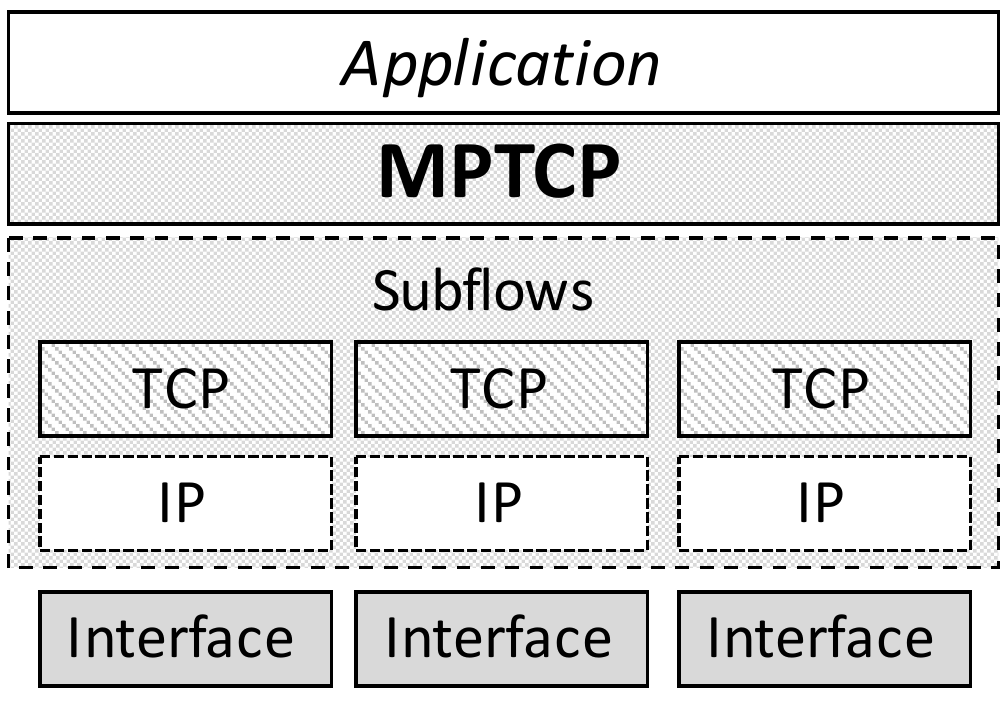}
        \caption{multihomed}
        \label{fig:arch1}
    \end{subfigure}
    \begin{subfigure}[b]{0.24\textwidth}
        \centering
        \includegraphics[width=\textwidth]{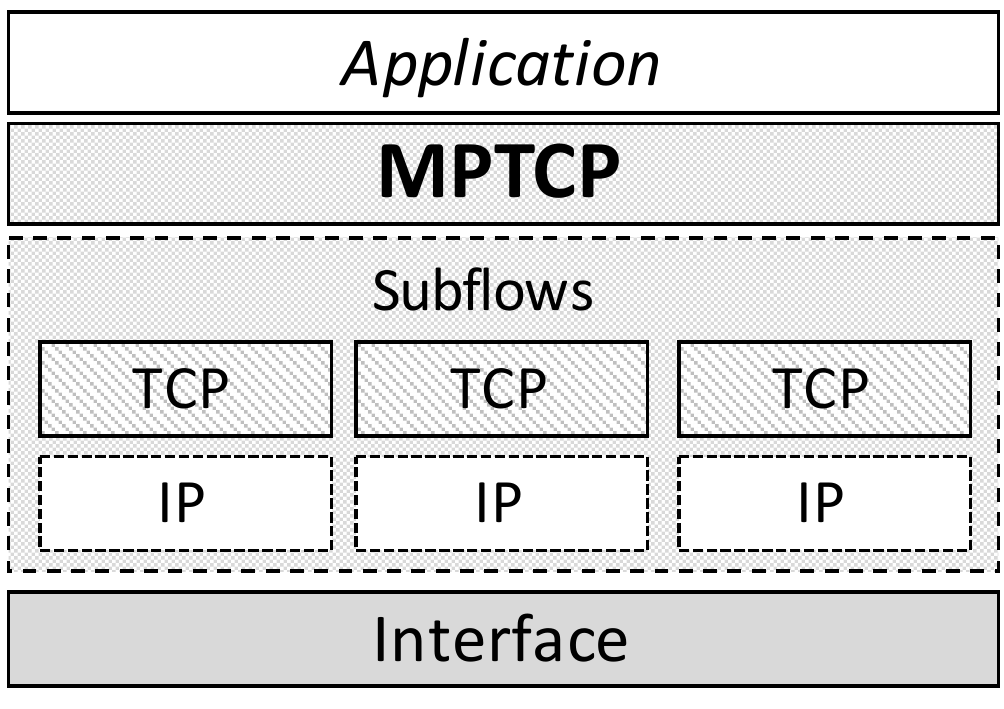}
        \caption{singlehomed}
        \label{fig:arch2}
    \end{subfigure}
    \caption{MPTCP Architecture}
    \label{fig:mptcparch}
\end{figure}

The Multipath TCP is an amendment of the TCP that enables concurrent data transmission over multiple paths~\cite{RFC6824}. As illustrated in Fig.~\ref{fig:arch1}, if we have a multihomed device, each interface can be associated to one TCP subflow. In singlehomed devices illustrated in Fig.~\ref{fig:arch2}, one single interface with multiple IP address can be used to map multiples subflows. The last exploits multiple possible routes in the network core to overcome issues such as congestion. Multipath TCP establishes a single end-to-end TCP connection and enables adding and removing paths (subflow TCP) while the connection is active. This protocol improves the performance offered by a single flow and makes the connection more resilient using concurrent flows (i.e. paths). 

Multipath TCP uses loss-based congestion control algorithms, such as TCP NewReno. Congestion control algorithms change only the Congestion Avoidance phase and maintain the Slow Start, Fast Retransmit and Fast Recovery phases as in the default TCP. Congestion control needs to satisfy the following goals: (1) improve throughput; (2) do not harm; (3) balance congestion. First, multipath subflows should perform as well as a single TCP flow through the best path. Second, the multipath subflow must be fair with TCP flow when they share a bottleneck. Third, multipath subflow should move the traffic to the less congested paths~\cite{li:16}. Multipath TCP congestion control executes at the subflow level, using a congestion window (CWND) by subflow and one shared receive window (RCVW) to ordered data delivery~\cite{zhou:15}.

Multipath TCP congestion control algorithms apply different approaches to AIMD (Additive Increase/Multiplicative Decrease) phase. The CWND of subflows are coupled in order to balance the traffic across various paths. It is a TCP modification that maintains a CWND on each active path. The CWND increases or decreases in response to acknowledgements and drops (loss) on that path, respectively. The increasing or decreasing factors of paths depend on CWNDs and RTTs from all other active paths. Raiciu et al.~\cite{raiciu:09} show that choosing appropriately the increasing and decreasing factor allows to achieve the congestion control goals cited up. Authors had proposed the Linked Increase Algorithm (\emph{LIA}) and the RTT Compensator. LIA has been adopted by IETF~\cite{RFC6824} as default Multipath TCP congestion control algorithm.

The LIA increasing factor rules are taken from the erstwhile congestion control algorithm, i.e. Fully Coupled in Eq.~\ref{eq:fully}, and the decreasing factor rules from Uncoupled TCPs, in Eq.~\ref{eq:uncoupled}~\cite{raiciu:09}. The $w_{i}$ parameter refers to the congestion window and $RTT_{i}$ to the RTT of a subflow $i$. The $W$ parameter regards to total value of congestion window for all subflows. 

In Eq.~\ref{eq:lia}, the choice for an appropriate value of $\alpha$ can limit the CWND and increase to be no more than a single TCP flow. This parameter, shown in Eq.~\ref{eq:alpha}, controls the subflows aggressiveness. Uncoupled TCP algorithm is different from others because it increases the CWND for all subflows in parallel (i.e. uncoupled), since $w_{i}$ increases with the same rate, i.e. one packet per RTT. However, this algorithm has difficulties in moving traffic away from a congested path. Thinking about heterogeneous scenarios, Raiciu et al.~\cite{raiciu:09} proposes RTT Compensator algorithm to compensate for dissimilarity of paths RTT. In Eq.~\ref{eq:rttc}, RTT Compensator takes the rules from LIA and Uncoupled TCP algorithms. Both algorithms LIA, Uncoupled and RTT\_Compensator use Eq.~\ref{eq:decrease} to CWND decreasing factor when receive a packet loss signal.

\begin{equation} \small \label{eq:uncoupled}
  w_{ i }=min(\frac { \alpha  }{ W } ,\frac { 1 }{ w_{ i } } )\quad (Uncoupled TCP)     
\end{equation}

\begin{equation} \small \label{eq:fully}
  w_{ i }=\frac { 1 }{ W } \quad (Fully Coupled) 
\end{equation}

\begin{equation} \small \label{eq:lia}
  w_{ i }=\frac { \alpha  }{ W } \quad (LIA)
\end{equation}

\begin{equation} \small \label{eq:rttc}
   w_{ i }=min(\frac { \alpha  }{ W } ,\frac { 1 }{ w_{ i } } ) \quad (Rtt\_ Compensator)
\end{equation}

\begin{equation} \small \label{eq:decrease}
  w_{ i }=\frac { w_{ i } }{ 2 } \quad (decrease factor)
\end{equation}

\begin{equation}  \small \label{eq:alpha}
        \alpha = 
        W * \frac{
          max\left(
              w_{i}/RTT_{i}^{2}
          \right)
        }{
        \left(
        \sum_i w_{i}/RTT_{i}
    \right)^{2}
    }
\end{equation}

Multipath TCP has as main goal to increase the performance of the transmission by using concurrent multipath transmission. However, congestion control algorithm that uses a coupled congestion window and a common receiver buffer turns the transmission performance dependency of the path diversity. Path diversity, such as delay variation and loss, seriously affects the  performance by increasing the occurrence of packet reordering and receiver buffer blocking~\cite{alfredsson13}. Many studies have shown that the differences between active path in heterogeneous wireless networks scenarios in terms of delay, caused by bufferbloat phenomenon, and losses, caused by dropping packets in full queues, figure as an important cause of performance degradation~\cite{ali:2015,ferlin14-2}. Congestion control algorithms need to deal with path diversity, in order to minimize packets arriving out-of-order and to maximize the use of link capacity~\cite{alheid:16}. In this work, we investigate how losses, caused by queue mechanism that control the queue size, affects the paths and whole concurrent multipath transmission. 

\section{Path delay management with CoDel}\label{sec:codel}

The performance of TCP-based applications, regardless of the congestion control mechanism, critically depends on the choice of queue management scheme implemented in routers. Queue management schemes control the size of the queues discarding proactively packets when it is necessary. The most common mechanism implements a PQM (Passive Queue Management) scheme, i.e. a queue FIFO discipline to forward packets. Whether the queue becomes full, PQM drops the latest incoming packets (DropTail). PQM scheme cannot deal with the problem of filling the large buffers, evidenced by Nichols and Jacobson \cite{nichols:12} as bufferbloat phenomenon.

With a congestion persistence, DropTail leads to wide variations in delay. This results in long queues that increase the RTT and degrade throughput, affecting the network performance. This shows limitations in PQM schemes. IETF 
recommends Active Queue Management (AQM) mechanisms for the next generation of Internet routers~\cite{rfc2309,rfc7567}. AQM mechanisms reduce significantly the latency across an Internet path, monitoring and limiting the growth of queues on the routers. These mechanisms try to reduce congestion proactively informing the end-hosts protocols that implement a loss-based congestion control through the packet drop. The early AQM disciplines used in routers, such as Random Early Detection (RED), depend on many parameter adjustments that vary according to the environment. Finding the optimal values for the parameters has been a critical problem, becoming reluctant to mass adoption of AQM~\cite{raghuvanshi:13}. Unlike RED, Nichols and Jacobson \cite{nichols:12} proposed a parameterless AQM Controlled Delay (CoDel). CoDel has a few adjustments, being easily deployed.

\begin{equation}
\small
    \label{eq:codel}
    \delta_{i} = deq_{i}-enq_{i}
\end{equation}

CoDel uses the packet sojourn time on queue as a metric to predict a congestion and control the queue size. CoDel algorithm follows two steps. In the first step, illustrated by Algorithm~\ref{al:codelenq}, on arrival of packet $p_{i}$, CoDel initially checks the current queue size. If queue has room, CoDel adds a timestamp (enqueue time) $enq_{i}$ in $p_{i}$ header and stores it.  

\begin{algorithm}
\small
\caption{CoDEL - Enqueuing packet}
\label{al:codelenq}
\begin{algorithmic}
  
  \STATE \textbf{On packet $p_{i}$ arrival:}
  \IF{Queue Size $<$ Queue Limit}
    \STATE Enqueue packet $p_{i}$
    \STATE $enq_{i}$ = timestamp for enqueue time
    \STATE Attach $enq_{i}$ in packet header
  \ELSE
    \STATE Drop packet $p_{i}$
  \ENDIF
  \STATE
 \end{algorithmic}
\end{algorithm}

\begin{figure}[!ht]
  \centering
  \includegraphics[width=0.3\textwidth]{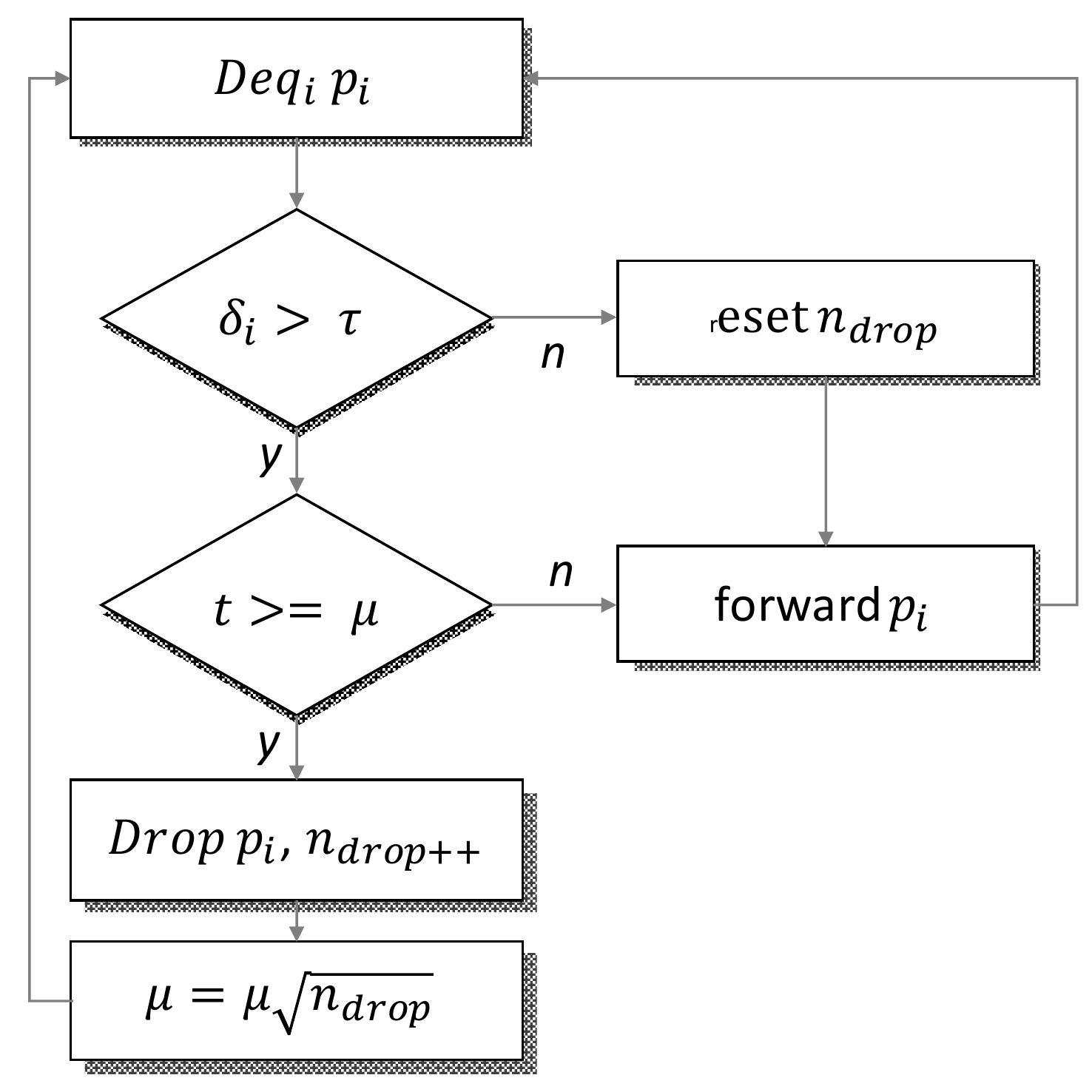}
  \caption{CoDel-FIFO}
  \label{fig:fifocodel}
\end{figure}

In the second step, illustrated in Fig.~\ref{fig:fifocodel}, when dequeuing the packet $p_{i}$ at time $deq_{i}$, CoDel calculates the delay or sojourn time of packet in queue (Eq.~\ref{eq:codel}). Sojourn time is used by CoDel to toggle between two states: dropping and non-dropping. The CoDel starts at non-dropping state and can change the status every $\mu$ interval. The initial $\mu$ value is $\lambda$ and it fits with the number of consecutive drops ($n_{drop}$). If a packet $p_{i}$ is dequeued between the time interval $t_{1} = t$ and $t2 = t + \mu$, such that $deq_{i} \in [t_{1},t_{2}]$ and $\delta_{i} > \tau$, CoDel enters in dropping state and the next dequeue packets are dropped while $ \delta_{i}> \tau $. Each consecutive packet drop increases $ n_{drop}+= 1$ and update $\mu$ value to $\mu\sqrt{n_{drop}}$. If $\delta_{i} < \tau$, CoDel enters in the non-dropping state, being the drop count ($n_{drop}$) reset to $1$ and $\mu$ to $\lambda$~\cite{kulatunga:15}.

The Flow Queue CoDel (FQ-CoDel) variation uses the same CoDel approach to control the packet sojourn time in queue. FQ-CoDel separates each subflows in one queue to apply the CoDel mechanism. Intending to provide fairness, packets are scheduled with a Deficit Round Robin (DDR) algorithm. The DDR algorithm sets a quantum of bytes ($Q_{i} $) to each $n_{i}$ queue in every round. If queue $n_{i}$ has no $Q_{i}$ bytes, this deficit accumulates to the next round. CoDel and FQ-CoDel suggest $\tau = 5ms$ and $ \lambda=100ms$ as the default value for most common situations~\cite{ali:2015}.

\section{CoDel-LIFO}\label{sec:lifocodel}

This section presents CoDel-LIFO, a new proposal to reduce the bufferbloat effects on multipath transmissions in heterogeneous wireless networks. 
The strategy lies in applying packet prioritization with a LIFO (Last In, First Out) queue discipline and in providing adjustments to reduce the number of packet drops, as detailed below. CoDel-LIFO is inspired by an adaptive strategy proposed in~\cite{maurer:15} to control the overload requests on Facebook servers. This work uses a CoDel algorithm and a LIFO discipline in the requests queue of server to prioritizing the newer request once that one oldest may have expired or renewed. Similarly, LIFO discipline intends to prioritize newer packets considering that the oldest has more chance to be dropped because it has a longer sojourn time.

\begin{figure}[!ht]
  \centering
  \includegraphics[width=0.28\textwidth]{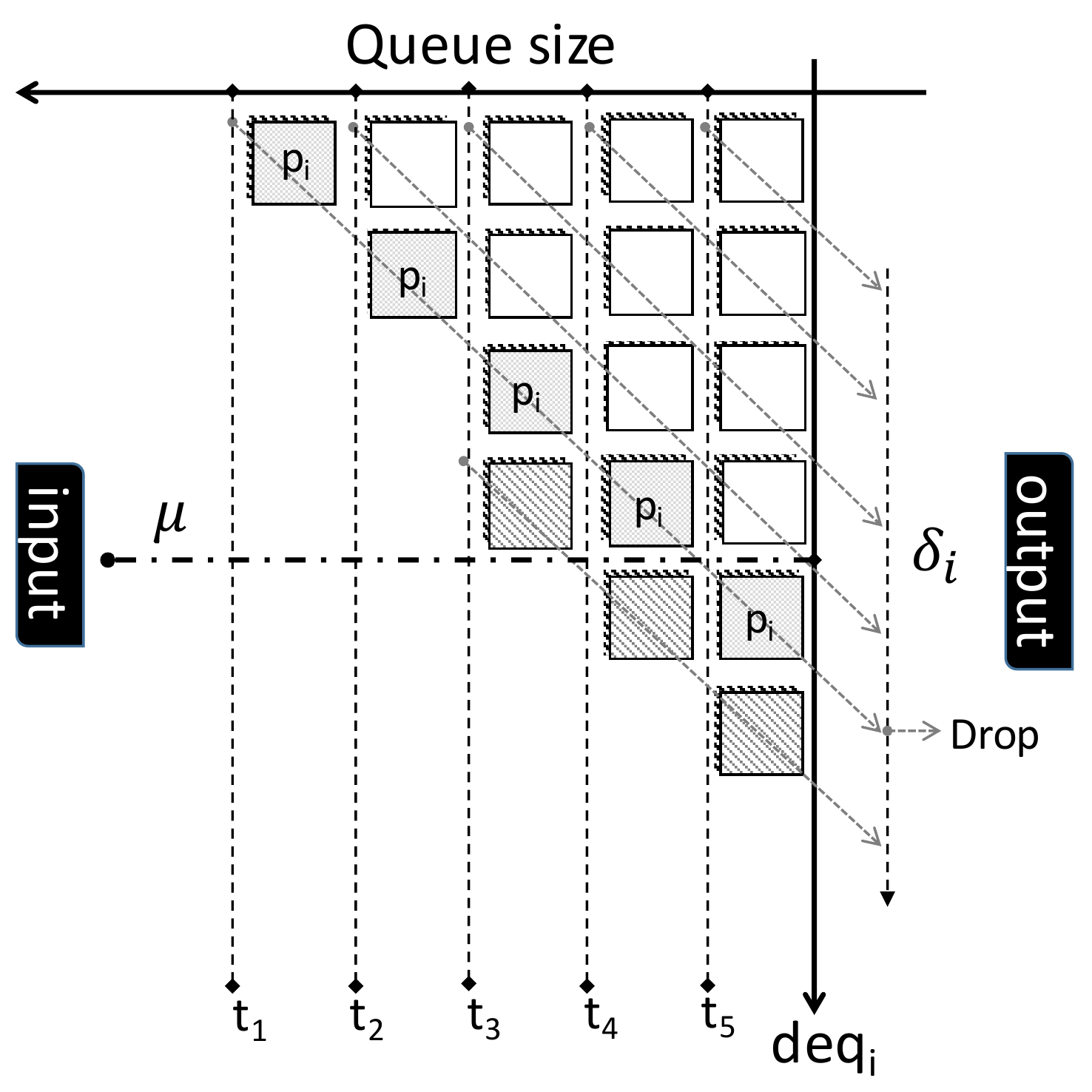}
  \caption{CoDel-FIFO queue}
  \label{fig:fifoc}
  \vspace{-0.3cm}
\end{figure}

Unlike the proposed approach, CoDel uses a FIFO (First In, First Out) mechanism to manage the arrival and departure of packets in queue. The Cartesian plane of the queue size $vs.$ dequeue, in Fig.~\ref{fig:fifoc}, illustrates the packets journey in CoDel-FIFO. Packet $p_{i}$ illustrates a drop example. The drop occurs when the sojourn time $\delta_{i}$ is greater than $\tau$ inside a dropping state (i.e. $deq_{i}$ is greater than $\mu$). Henceforth we will use interchangeably CODEL and CODEL-FIFO.

As the queue grows, it increases the sojourn time of packets. However, if the queue shrinks, the sojourn time decreases. Regardless of the queue size, the packets withdrawn from the queue follows the order of arrival. This occurs because FIFO discipline deploys a queue structure to ensure that the input order is the same output order. CoDel uses the parameter $\tau$ to control the packet delay based in a fixed value. The parameter $\mu$, with initial value equals $\lambda$, controls the queue growth adjusting dynamically the drop state interval. The efficient use of transmission channel depends on achieving an optimal adjustment between the initial values of $\mu$ and $ \lambda$. In Section~\ref{sec:evaluation}, we show that parameters are not effective to multipath transmission. The mechanism causes many drops in queue and it impacts congestion control. Hence, the transfer rate is not sufficient to occupy the entire available channel.

In order to overcome this issue, we look for a solution to reduce the queue drop. Once CoDel has the best results against the bufferbloat and its logic is simple, we propose to add a self-adjustable parameter to control drops. However, we show that the queue structure does not permit an effective control of sojourn time. Analyzing the queue in Fig.~\ref{fig:fifoc}, as the queue size varies, the packets sojourn time also vary. For example, the packet enqueued at instant $t_{3}$ has a sojourn time shorter than the previous packet $p_{i}$, being removed at a later time.

\begin{figure}[!ht]
  \centering
  \includegraphics[width=0.28\textwidth]{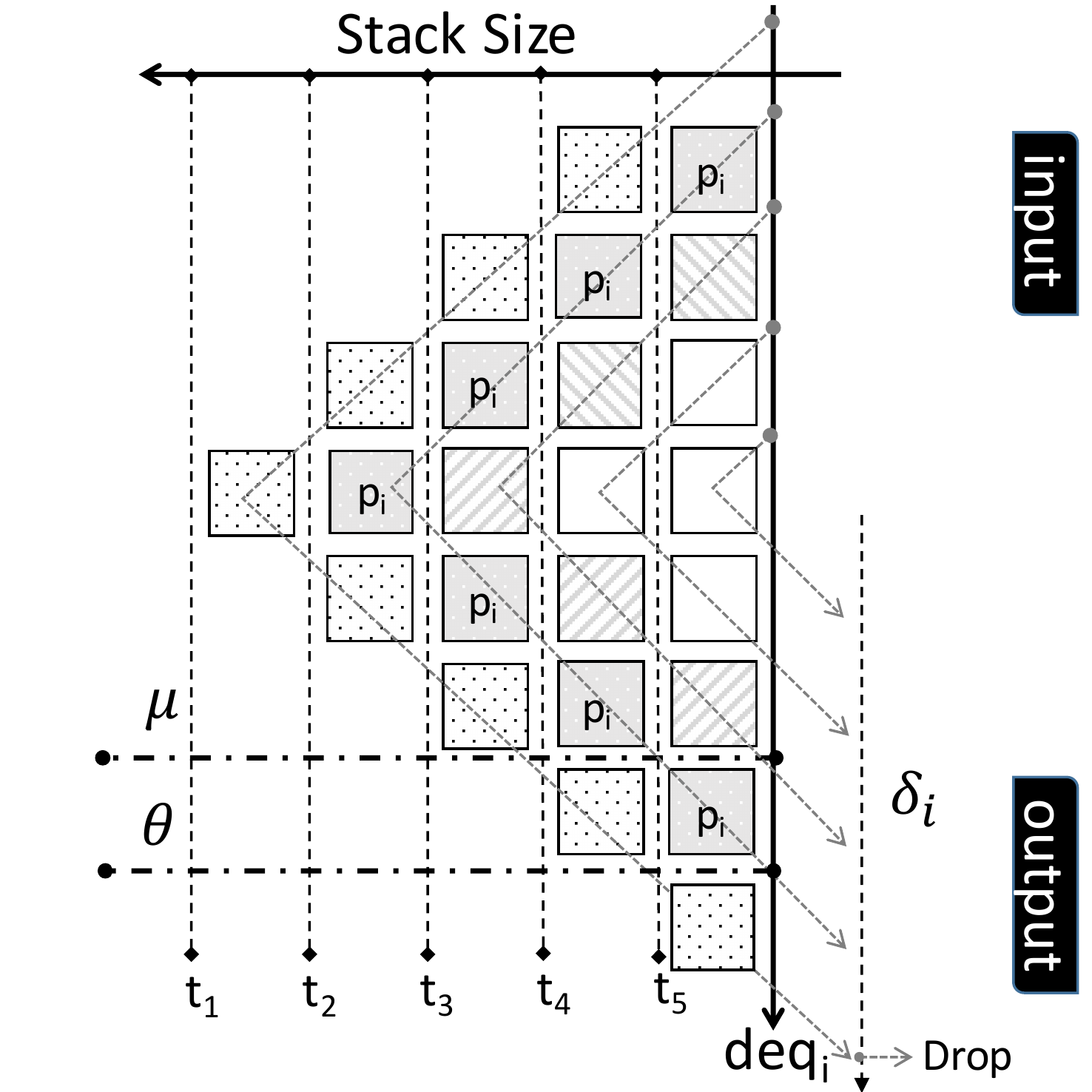}
  \caption{CoDel-LIFO stack}
  \label{fig:lifoc}
  \vspace{-0.5cm}
\end{figure}

Changing the queue to a stack structure (Fig.~\ref{fig:lifoc}), the packets are stacked and unstacked only on one side. Henceforth, we will use interchangeably stack to refer a LIFO queue. If one takes a snapshot of stack at an instant $t_{i}$, the sojourn time increases as the packets are removed. This structure also enables packet prioritization, enabling to forward the packets stored in the queue with shorter sojourn time. As the stack grows, packets near the base (i.e. longest time in stack) remain there for longer and have more chances to be dropped. The packets that remain longer in stack have a lower sequence number than others that are stacked and unstacked in this time interval. The packet $p_{i}$ has a lower sequence number than the upper packets. Hence, packets arrive in receiver out-of-order. This has a negative and a positive side. The delay caused by packet reordering in receiver is a negative aspect. The positive is that out-of-order packet triggers duplicate acknowledgements (dupACK) inducing the sender to infer a loss and then triggering congestion control.

\begin{figure}[!ht]
  \centering
  \includegraphics[width=0.28\textwidth]{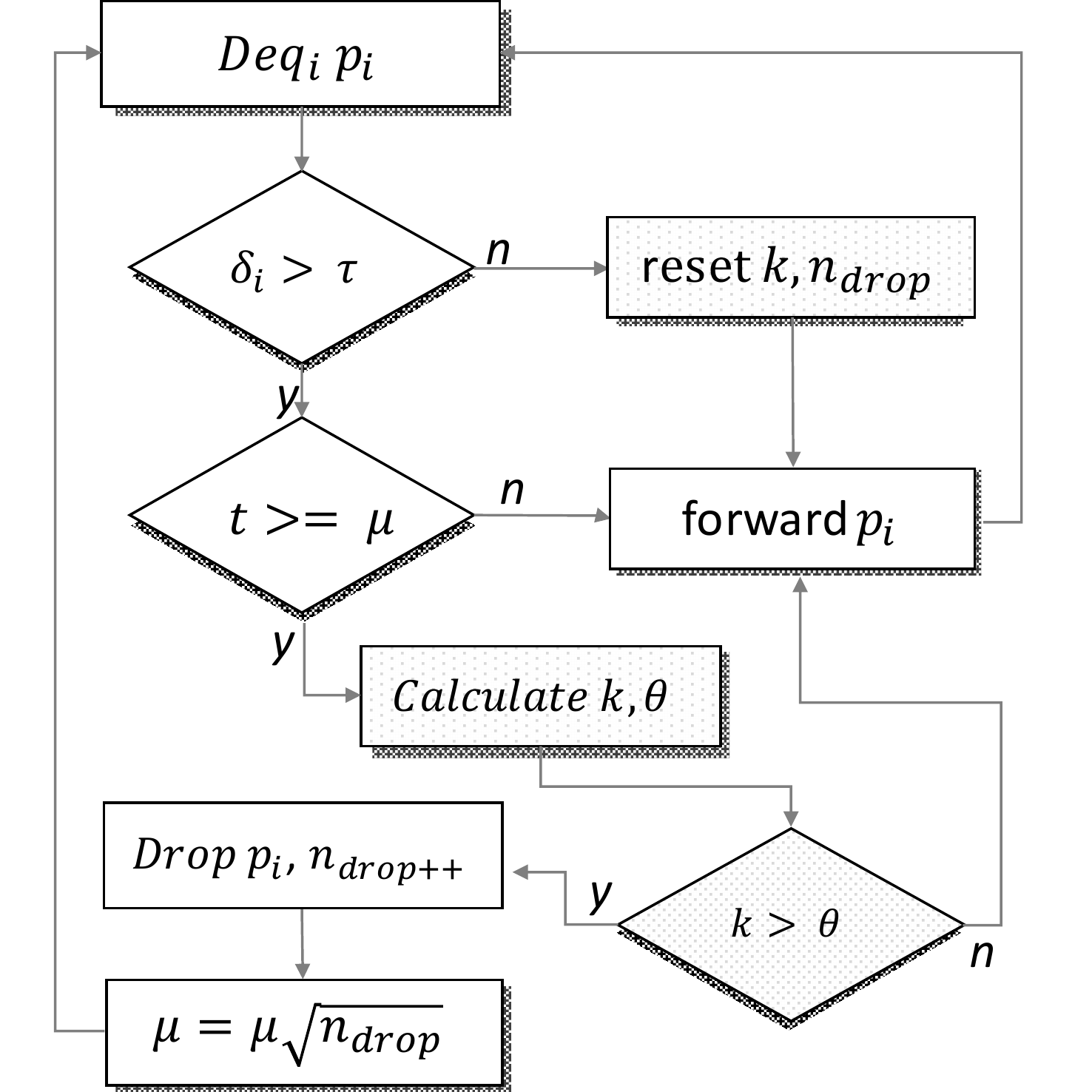}
  \caption{CoDel-LIFO}
  \label{fig:lifof}
  \vspace{-0.3cm}
\end{figure}

In addition to the stack advantages, we defined a forgiveness for doomed packets, i.e. forward the packet that would be dropped, in order to reduce the number of discards. With CoDel algorithm the packet $p_{i}$, in Fig.~\ref{fig:lifoc}, is dropped because it has a sojourn time greater than $\tau$ and $deq_{i} > \mu$ (i.e. above dropping state). To forgive this packet, CoDel-LIFO uses the $\theta$ parameter in dequeue algorithm, illustrating in the flowchart of the Fig.~\ref{fig:lifof}. After checking sojourn time and dropping state, $p_{i}$ will be dropped only if the variable $k > \theta$.

The parameter $\theta$ (Eq.~\ref{eq:theta}) represents the ratio between the maximum and mean sojourn time calculated every unstacked packet. The $\delta_{max}$ value refers to the maximum sojourn time of the unstacked packets from the stack, as Eq.~\ref{eq:delta_max}. The $\bar{\delta}$ represents the average sojourn time for the \emph{n-th} packet unstacked (Eq.~\ref{eq:delta_av}). If the difference between the $\delta_{max}$ and the $\bar{\delta}$ is small and the number of samples is large, $\theta$ value approximates to $2$. The $\theta$ value serves as an upper limit used to control sending doomed packets. However, $k$ counts the number of doomed packet unstacked with crescent sojourn time. CoDel-LIFO calculates the $\Gamma_{t}$ value before $k$. $\Gamma_{t}$ refers to the difference between earlier and current sojourn time (Eq. \ref{eq:gamma}). Whenever $\Gamma_{t} > 0$, the $k$ variable is incremented. Otherwise, $k$ receives $0$ (Eq. \ref{eq:kvalue}). $\Gamma_{t} <= 0$ indicates that unstacked packet is newer than the earlier, but the sojourn time can be bigger than $\tau$. All variables are initialized when the sojourn time is less than $\tau$.
\begin{equation} 
\small
    \label{eq:delta_max}
    \delta_{max}=max(\delta_{1},\ldots,\delta_{n})
\end{equation}

\begin{equation} 
\small
    \label{eq:delta_av}
    \bar{\delta}=\frac{1}{n}\sum_{i=1}^{n}{\delta_{i}}
\end{equation}

\begin{equation} 
\small
    \label{eq:theta}
    \theta=\frac{{\delta}_{max}}{\bar{\delta}},\bar{\delta}\neq 0
\end{equation}

\begin{equation}
\small
    \label{eq:gamma}
    \Gamma_{t}=\delta_{i}-\delta_{i-1}
\end{equation}

\begin{equation}
\small
    \label{eq:kvalue}
    k=\begin{cases}k+1,\quad if\quad \Gamma_{t}>0\\0,\quad if\quad \Gamma_{t}\le 0\end{cases}\\
\end{equation}

The parameters $\theta$ and $k$ are self- adjusted. If the value of $k$ grows the indication of the draining of the stack. Otherwise, $\theta$ can only be a high value if the difference between the max value and mean value is high. This only occurs if the longest time packet in the stack is unstacked. Hence, $\theta - k$ packets are sent until the next drop. The goal is to reduce the drops and to enable forwarding packets with long time on the stack that CoDel algorithm would drop these packets. CoDel-LIFO purpose is reduce the drops and the sojourn time of packets in stack. With less drops, CoDel-LIFO has less impact in Multipath TCP congestion control. These changes can have a negative impact on packet reordering on receive buffer. However, the early detection of congestion with dupACKs makes this drawback to have a minor impact.


\section{Evaluation}\label{sec:evaluation}

In order to compare the impact of CoDel-LIFO to CoDel and DropTail over Multipath TCP, NS-3 simulations~\cite{morteza2015} consider three congestion control algorithms, i.e. LIA, RTT Compensator and Uncoupled TCP. Each simulation scenario employs a combination of a queue discipline and a congestion control algorithm. This section proceeds as follows. Subsection~\ref{subsec:scenario} overviews the baseline scenario, parameters and factors; Subsection~\ref{subsec:metrics} describes the metrics to evaluate congestion control algorithms and queue discipline performance. Subsection~\ref{subsec:fiforesults} shows results comparing CoDel and DropTail with congestion control algorithms, highlighting the trade-off between latency and goodput; Subsection~\label{subsec:liforesults} shows results comparing the proposed solution and CoDel, emphasizing how CoDel-LIFO reduces the latency and goodput trade-off.



\subsection{Simulation Setup}\label{subsec:scenario}
\begin{figure}[!ht]
\centering
\includegraphics[width=3.2in]{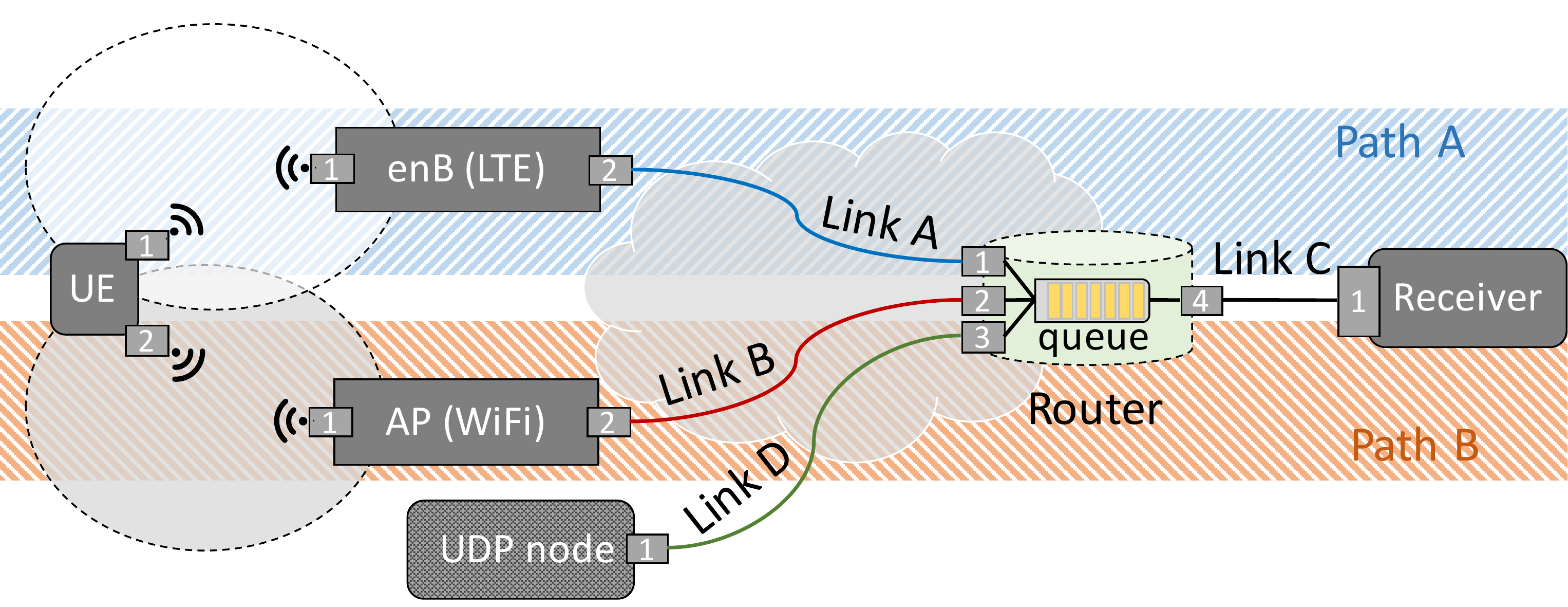}
\caption{HetNet topology in the baseline simulation scenario}
\label{fig:scenario}
\end{figure}

Fig.~\ref{fig:scenario} illustrates the baseline scenario employed for comparisons. This scenario represents a common heterogeneous multipath transfer environment to multihomed devices (i.e. smartphone or tablet). It is broadly employed in recent works~\cite{ferlin14-2,ferlin14,chen14}. It comprises of two heterogeneous access networks and a multihomed user equipment (UE) that send data for a receiver node. The two access networks follow LTE and WiFi technologies. This scenario is complemented with a router (shared queue) between end-nodes to evaluate queue aspects. The router has the role of a border gateway and provides access between access networks LTE (link A) and WiFi (link B), UDP node (link D), with receiver (link C). UE can use concurrently the path A (through LTE) and path B (through WiFi) to send TCP packets to receiver.


 Packets traveling through path A and B share the same router queue and the bottleneck, i.e. {\em link C} between the router and receiver. Multipath TCP uses slow start algorithm to control the send rate growing and it slowly fills the queue. UDP node generates a workload (UDP packets) and sends to receiver with a constant bit rate ($0.25$ Mbps) to provide a rapidly queue filling. For all simulations, UE establishes a Multipath TCP connection with receiver by LTE interface. After concluding the three-way handshake, UE opens a second connection through the WiFi. The subflows are linked to a single Multipath TCP connection, transmitting data concurrently.

\begin{table}[h]
\centering
\caption{Simulation parameters}
\label{table:factors}
\begin{tabular}{ll}
\hline
\textbf{Parameter}  & \textbf{Value} \\ \hline
Data Rate           & 1.0          \\
Delay - Path $A$       & 1, 10, 100, 300 ms\\
Queue Size      &     100 packets           \\
Packet Size      &     1458 bytes           \\
CWND, RCWD      &     64 Kb           \\
Congestion control     &     LIA, RTT\_Compensator, Uncoupled           \\
\hline
\end{tabular}
\end{table}


Simulations are performed with a fixed workload of $4$ Mbytes, i.e. for each run. UE transfers this amount of data to receiver simultaneously by paths A and B. Initially, links $A$, $B$ and $D$ have $1$ Gbps of bandwidth each and $1$ ms of delay. Link delay is fixed to provide controlled simulation scenario. However, real environments, especially in wireless networks, delays change constantly. Delay variation is represented in simulation varying link A delay from $1$ to $300$ ms. 

Table~\ref{table:factors} shows the parameters and factors employed in simulations. The bottleneck (link C) has $1$ Mbps of bandwidth to force a bufferbloat in shared queue. Packet size is defined with $1458$ bytes to auxiliary the queue rapid fill. Both congestion window and receive window use default parameter values of NS-3. The access point (WiFi standard IEEE802.11a) and enB (LTE) uses default simulator parameters for the MAC and PHY layers. Increases of delay in link $A$ provides a RTT variation between path A and B that represents a heterogeneous wireless network environment with a mobile node, as in Lee and Lee~\cite{lee:15}. Delay difference forces some congestion control algorithms to give preference for the fast path (low RTT).

\subsection{Metrics}\label{subsec:metrics}

The following metrics have been employed to assess the Multipath TCP congestion control algorithms and queue disciplines performance: $(i)$ Average goodput (Eq.~\ref{eq:goodput_metric}); $(ii)$ average RTT per path (Eq. \ref{eq:rtt_metric}); (iii) average number of dropped packets (Eq.~\ref{eq:drp_metric}); and $(iv)$ average queue length and average sojourn time (Eq.~\ref{eq:qt_metric}). In the equations, $n$ refers to number of repetitions (i.e. seed changes) and $t$ to the simulation time taken to transfer the workload in each run. Sojourn time and queue length are calculated only for CoDel-based disciplines.

\begin{equation} \small
\label{eq:goodput_metric}
  \begin{aligned}
      GoodPut&=\sum{\overline{GdpPath_{k}}} \\
      \overline{GdpPath_{k}}&=\cfrac{1}{n}\sum_{i=1}^{n}{Gdp_{i}} \\
      Gdp_{i}&=\frac{bytes\_received\quad \times \quad 8}{TimeLastPkt-TimeFirstPkt}
  \end{aligned}
\end{equation}
\newline
\begin{equation} 
\small
    \label{eq:rtt_metric}
     \begin{aligned}
         \overline{{RttPath}_{k}}&=\cfrac{1}{n}\sum_{i=1}^{n}{{RttRep}_{i}}\\
         RttRep_{i}&=\cfrac{1}{t}\sum_{j=1}^{t}{{Rtt}_{j}}  
     \end{aligned}
\end{equation}
\newline
\begin{equation} 
\small
    \label{eq:drp_metric}
    \begin{aligned}
      \overline{DropsQueue}&=\frac{1}{n}\sum_{i=1}^{n}{DropRep_{i}}\\
      DropRep_{i}&=\sum{Drop} 
    \end{aligned}
\end{equation}

\begin{equation} 
\small
    \label{eq:qt_metric}
    \begin{aligned}
      \overline{AverageValue}&=\frac{1}{n}\sum_{i=1}^{n}{Rep_{i}}\\
      Rep_{i}&=\frac{1}{t}\sum_{j=1}^{t}{V_{j}} 
    \end{aligned}
\end{equation}

\subsection{CoDel Impact on Congestion Control over Multipath TCP}\label{subsec:fiforesults}

CoDel controls the queue size discarding packet on queue. Although reduces drastically the packet queue delay, drops can have impact on congestion control algorithms loss-based and reduce the goodput. In order to asses this impact on multipath transmission, simulations are performed with CoDel and DropTail. This last serves as baseline because it is the default queue manager in most routers. Simulation results are presented with an average of $35$ repetitions and $95\%$ of confidence interval. The chart axis $x$ presents delay on link A varying from $1ms$ to $300ms$.

\begin{figure}[ht]
\vspace{-0.3cm}
    \centering
    \begin{subfigure}[b]{0.24\textwidth}
        \centering
        \includegraphics[width=\textwidth]{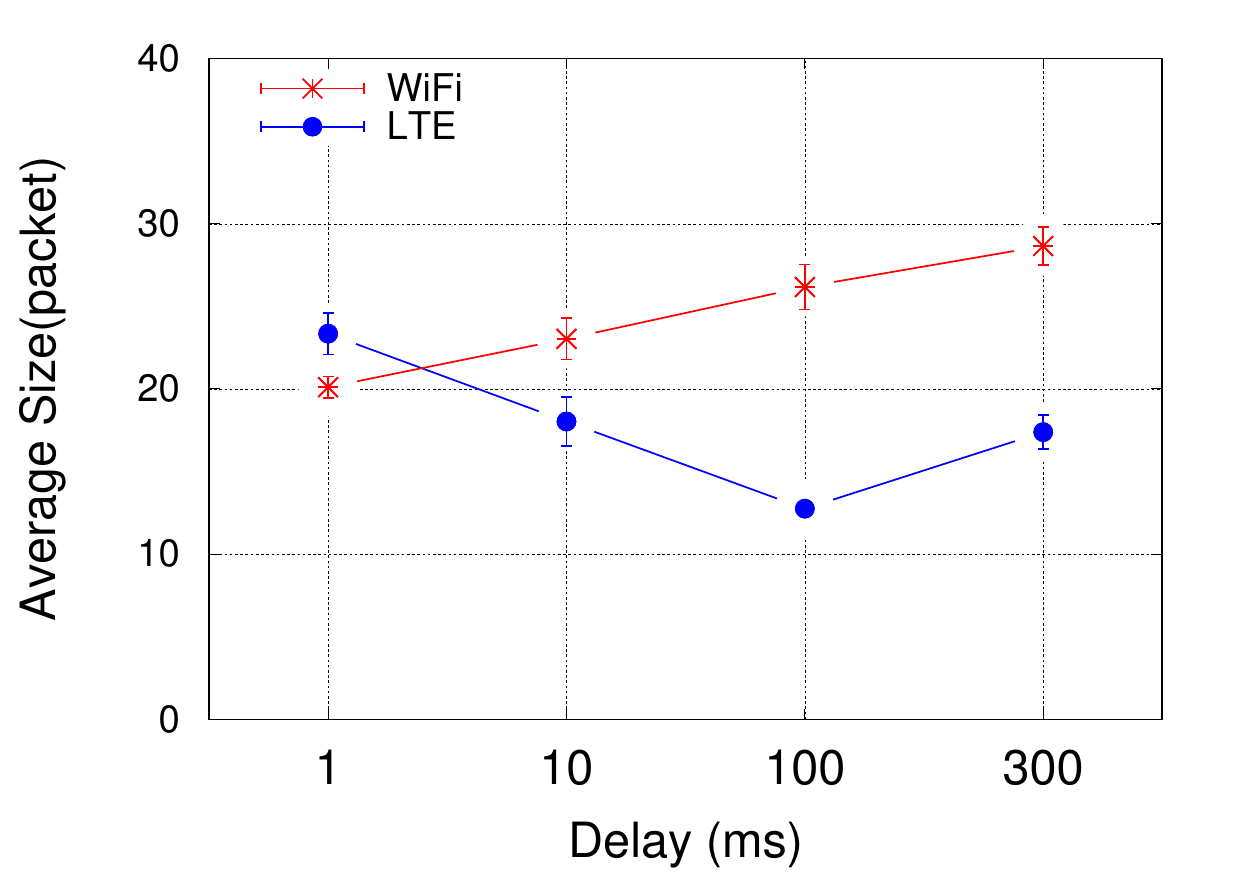}
        \caption{DropTail}
        \label{fig:droptailcwnd}
    \end{subfigure}
    \begin{subfigure}[b]{0.24\textwidth}
        \centering
        \includegraphics[width=\textwidth]{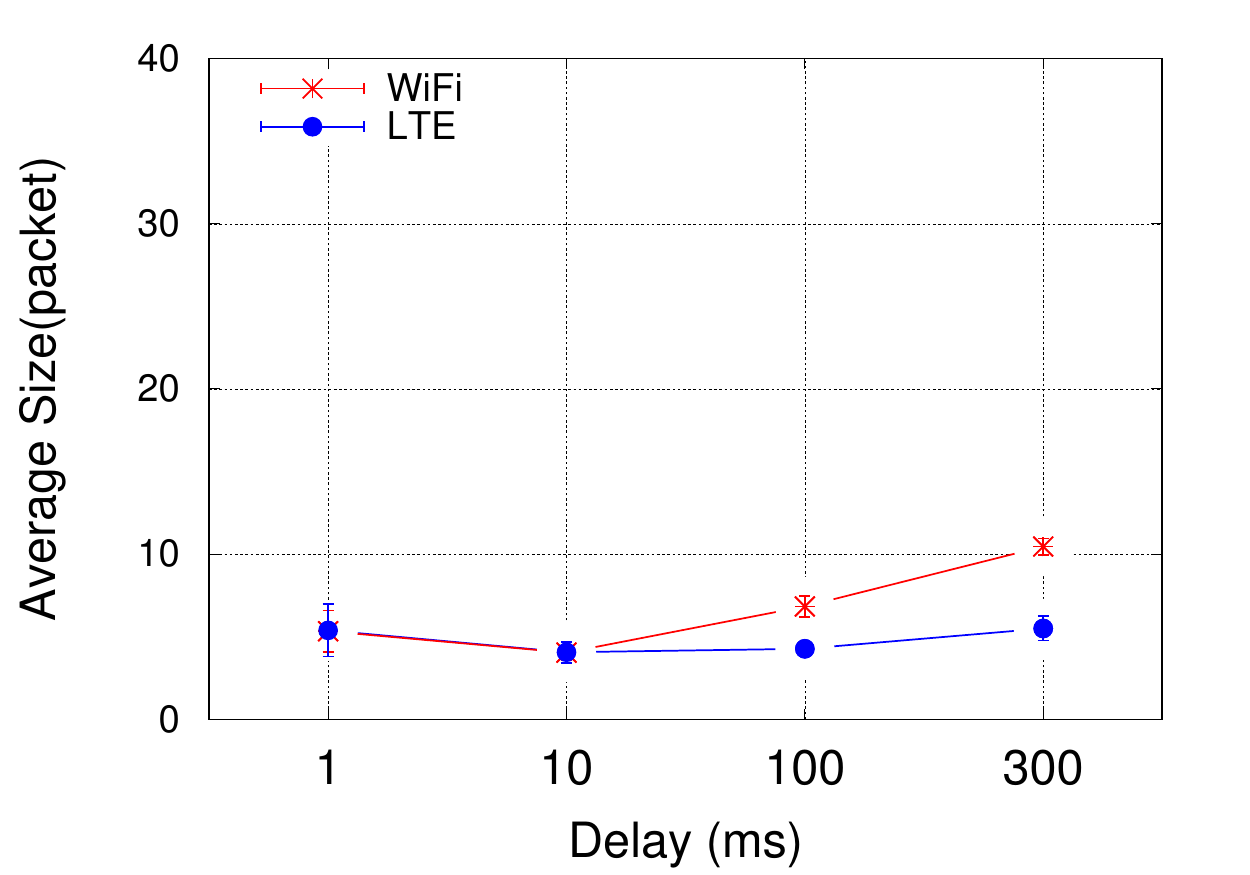}
        \caption{CoDel}
        \label{fig:fifocwnd}
    \end{subfigure}
    
\caption{CWND Results}
\label{fig:codelcwnd}
\vspace{-0.3cm}
\end{figure}

Fig.~\ref{fig:droptailcwnd} shows the average CWND size with LIA and DropTail. DropTail discards packets only when the queue is full. Hence the CWND grows until the queue fills up. LIA prioritizes path B, that has lower RTT. However it does not happen properly with CoDel-based solutions. CoDel-based solutions prevent queue filling dropping packets immediately that the sojourn time is over the limit. The CWND of subflows remains very low.

\begin{figure}[!b]
\vspace{-0.5cm}
    \centering
    \begin{subfigure}[b]{0.24\textwidth}
        \centering
        \includegraphics[width=\textwidth]{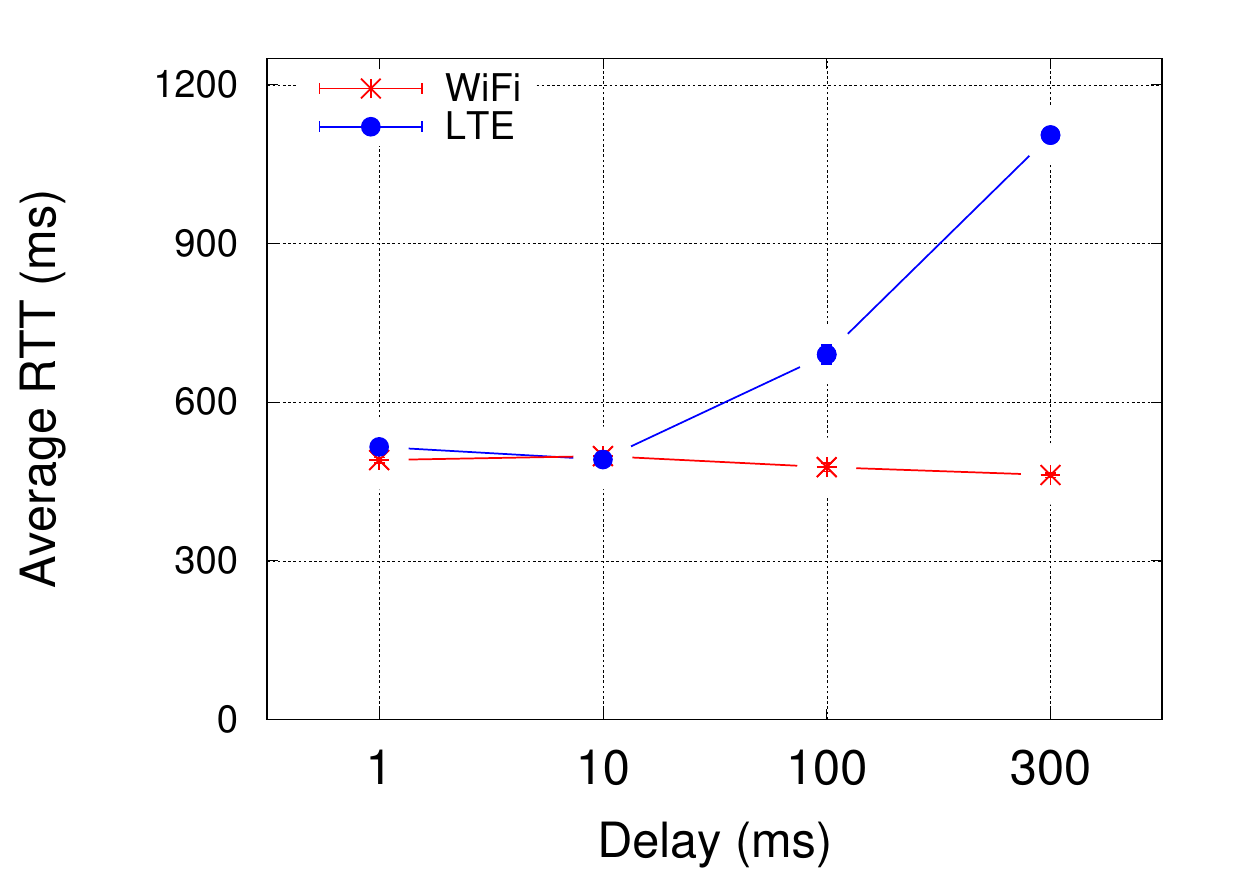}
        \caption{LIA vs. DropTail}
        \label{fig:liadroptail}
    \end{subfigure}
    \begin{subfigure}[b]{0.24\textwidth}
        \centering
        \includegraphics[width=\textwidth]{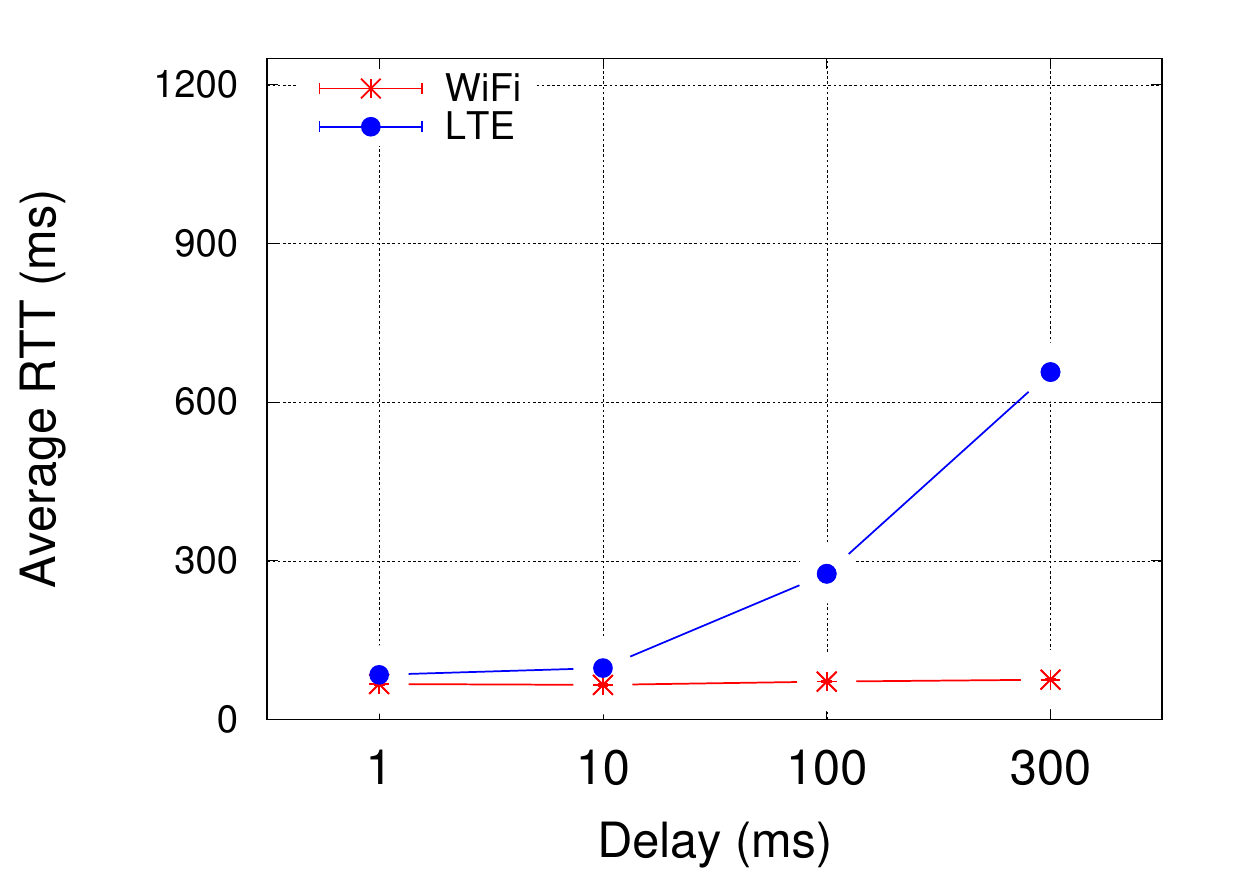}
        \caption{LIA vs. CoDel}
        \label{fig:rttccodel}
    \end{subfigure}
    
    \begin{subfigure}[b]{0.24\textwidth}
        \centering
        \includegraphics[width=\textwidth]{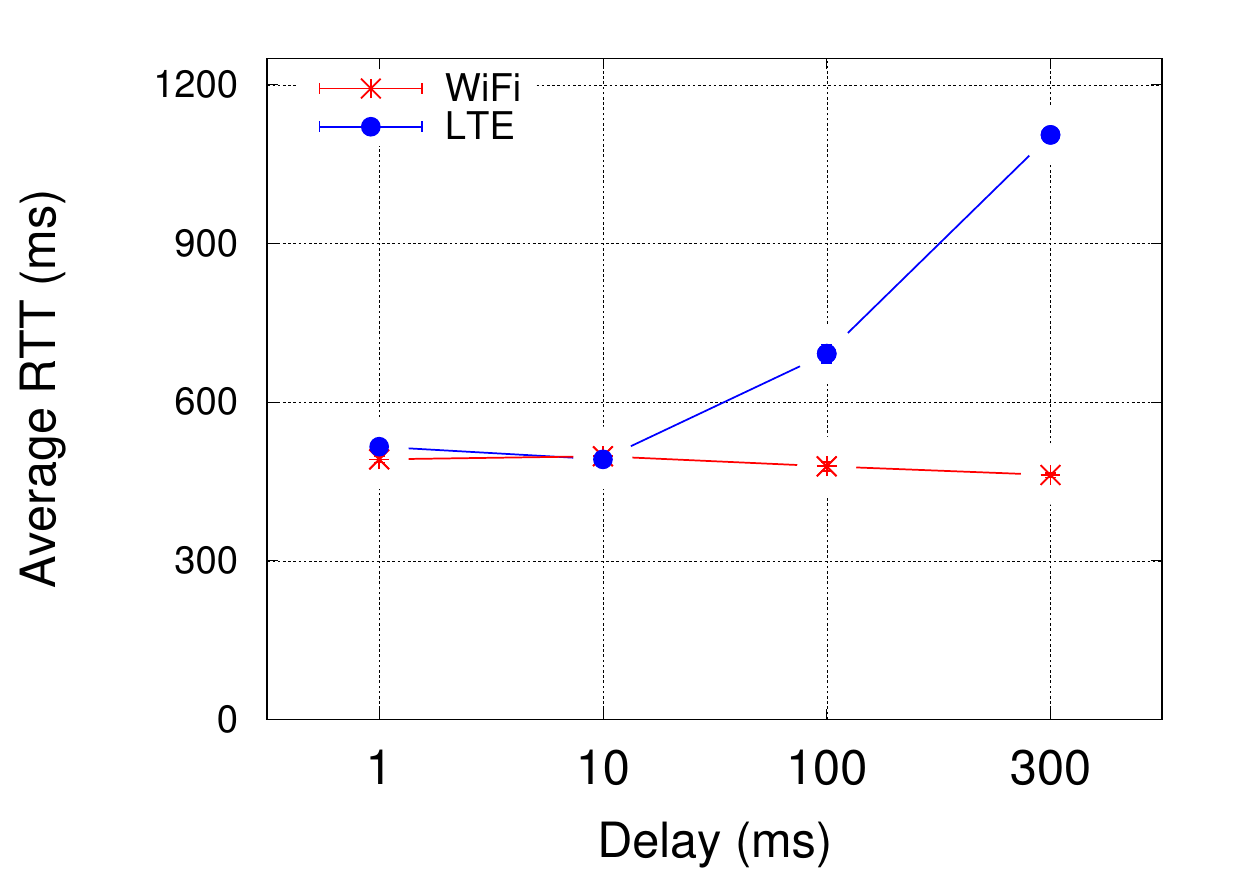}
        \caption{RTT Comp. vs. DropTail}
        \label{fig:rttccodel}
    \end{subfigure}
    \begin{subfigure}[b]{0.24\textwidth}
        \centering
        \includegraphics[width=\textwidth]{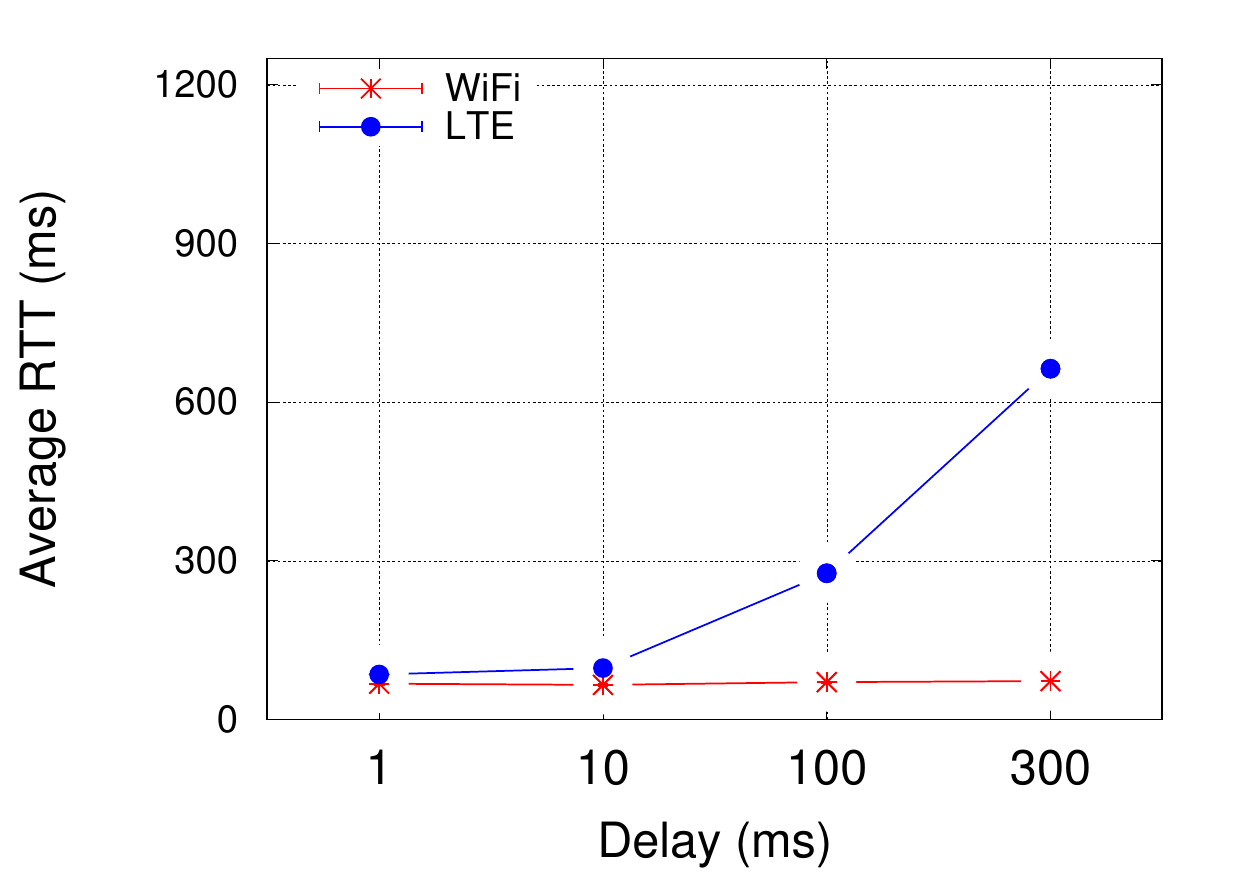}
        \caption{RTT Comp. vs. CoDel}
        \label{fig:liacodel}
    \end{subfigure}
    
      \begin{subfigure}[b]{0.24\textwidth}
        \centering
        \includegraphics[width=\textwidth]{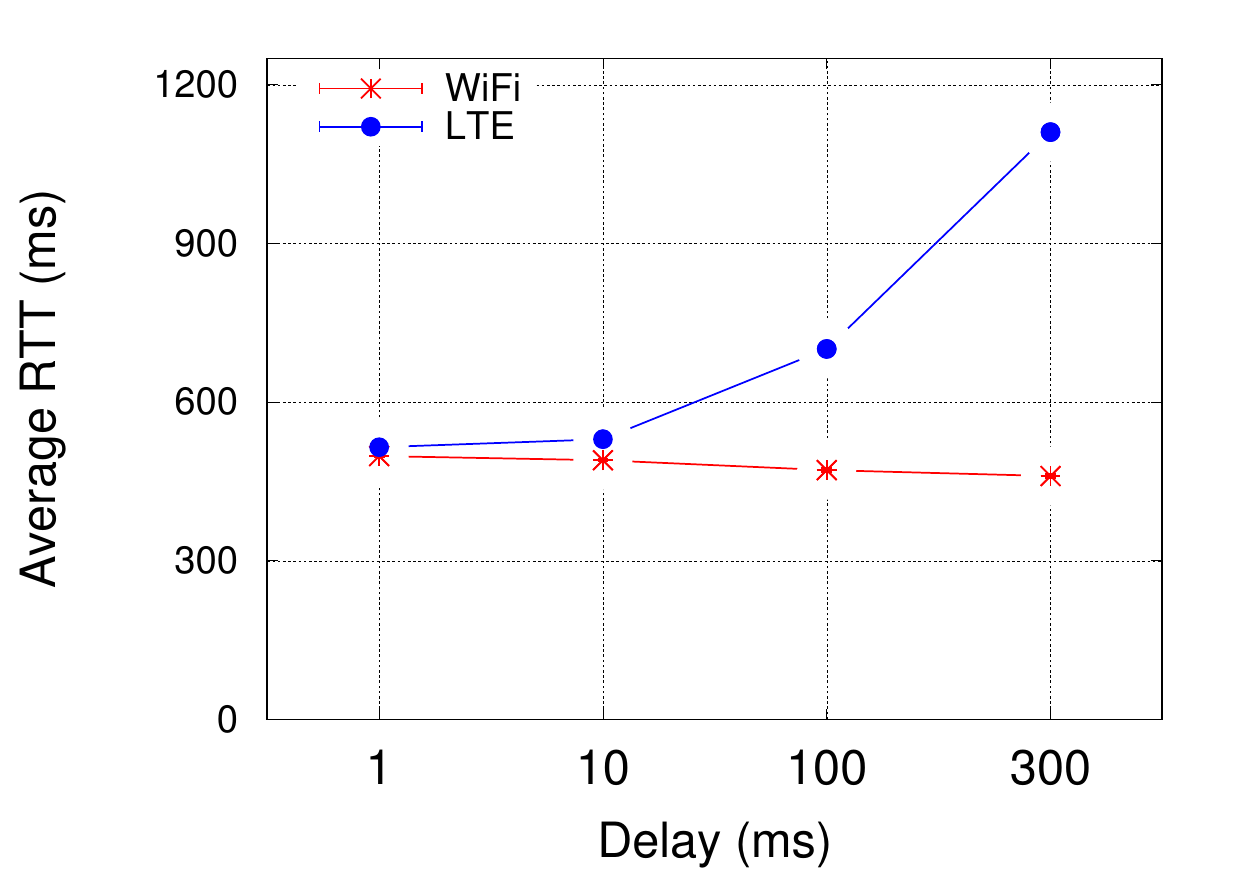}
        \caption{Uncoupled vs. DropTail}
        \label{fig:uncoupleddroptail}
    \end{subfigure}
     \begin{subfigure}[b]{0.24\textwidth}
        \centering        \includegraphics[width=\textwidth]{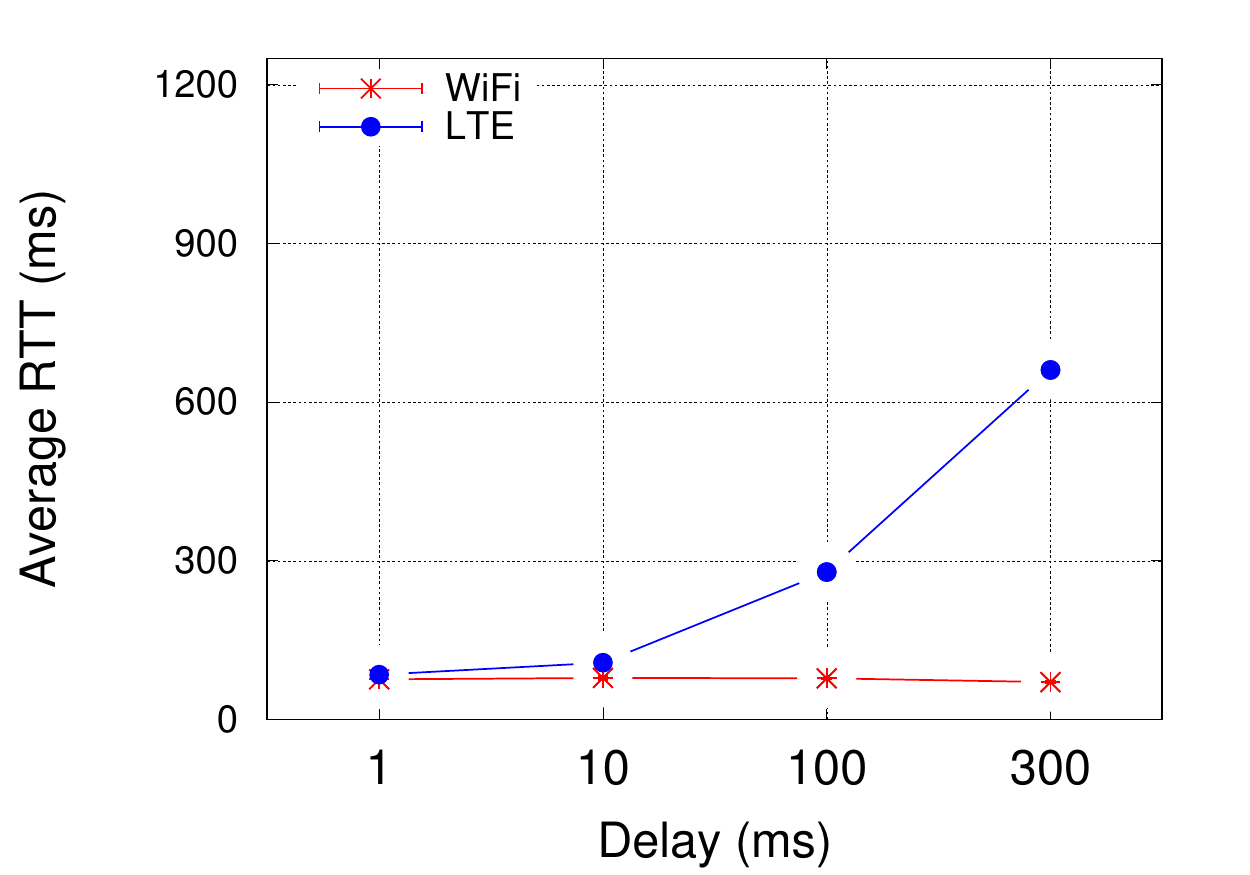}
        \caption{Uncoupled vs. CoDel}
        \label{fig:uncoupledcodel}
    \end{subfigure}
  
    \caption{RTT results}
    \label{fig:codel-rtt-impact}
    \vspace{-0.5cm}
\end{figure}

Fig.~\ref{fig:codel-rtt-impact} presents results for the average RTT metric. Plots show that queue discipline impact on congestion control algorithms. Confidence interval bars do not appear in the plots because the max error does not exceed $15ms$ and y-axis has a high scale. DropTail presents high RTT in comparison to CoDel over all congestion control algorithms. When both paths has the delay ($1ms$), DropTail presents $515ms$ against $84ms$ of the CoDel. As delay increases in path A, the difference in results for CoDel and DropTail gets higher. With delay of $300ms$ on path A and DropTail, RTT is closer to $1100$ms. Using CoDel, RTT is closer to $700ms$. RTT on path B have a slightly variation, but with CoDel is $8$ times less than DropTail. Related to congestion control algorithms, RTT Compensator has a RTT very close to those observed with LIA. Uncoupled algorithm with DropTail have a RTT slightly shorter and with CoDel slightly high.

\begin{figure}[ht]
    \centering
    \begin{subfigure}[b]{0.24\textwidth}
        \centering
        \includegraphics[width=\textwidth]{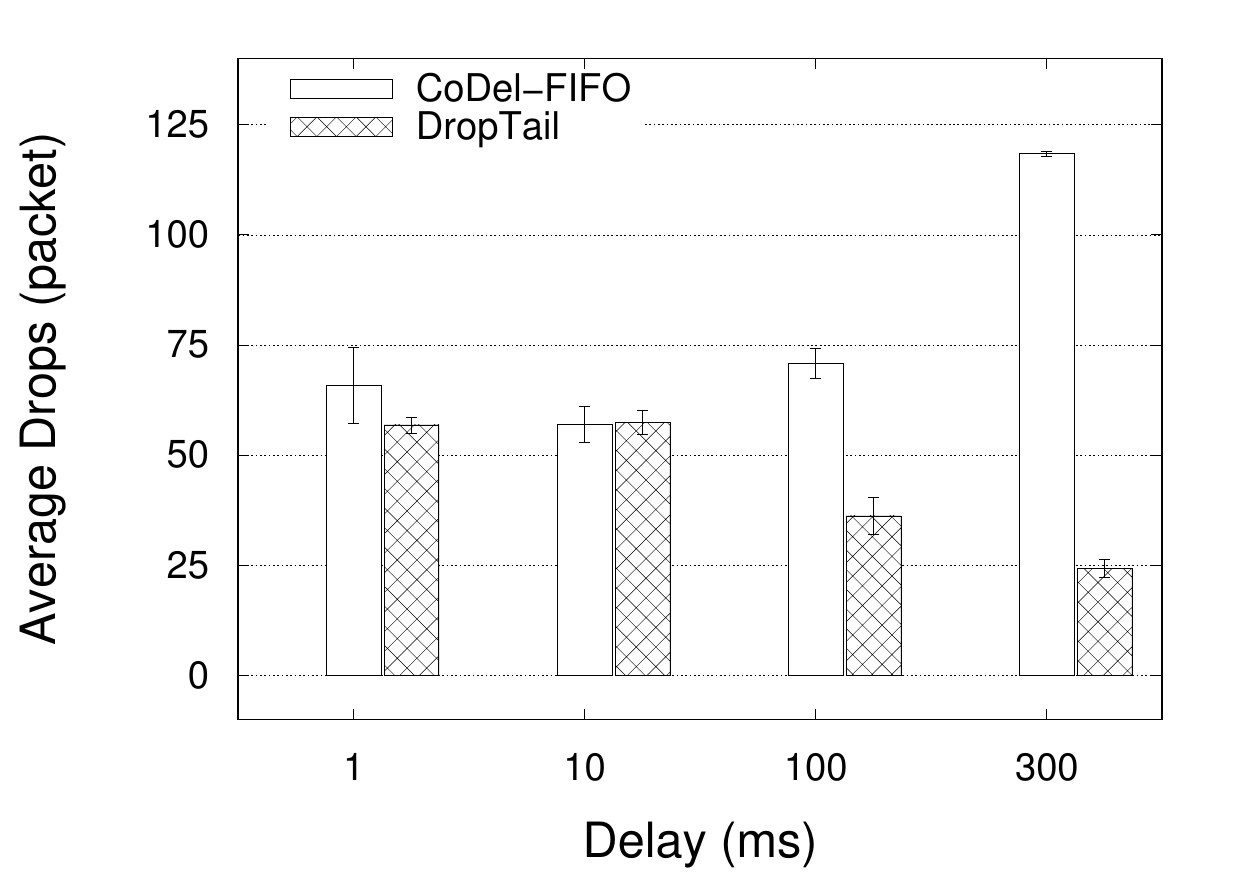}
        \caption{Queue Drops}
        \label{fig:queuedrops}
    \end{subfigure}
    \begin{subfigure}[b]{0.24\textwidth}
        \centering
        \includegraphics[width=\textwidth]{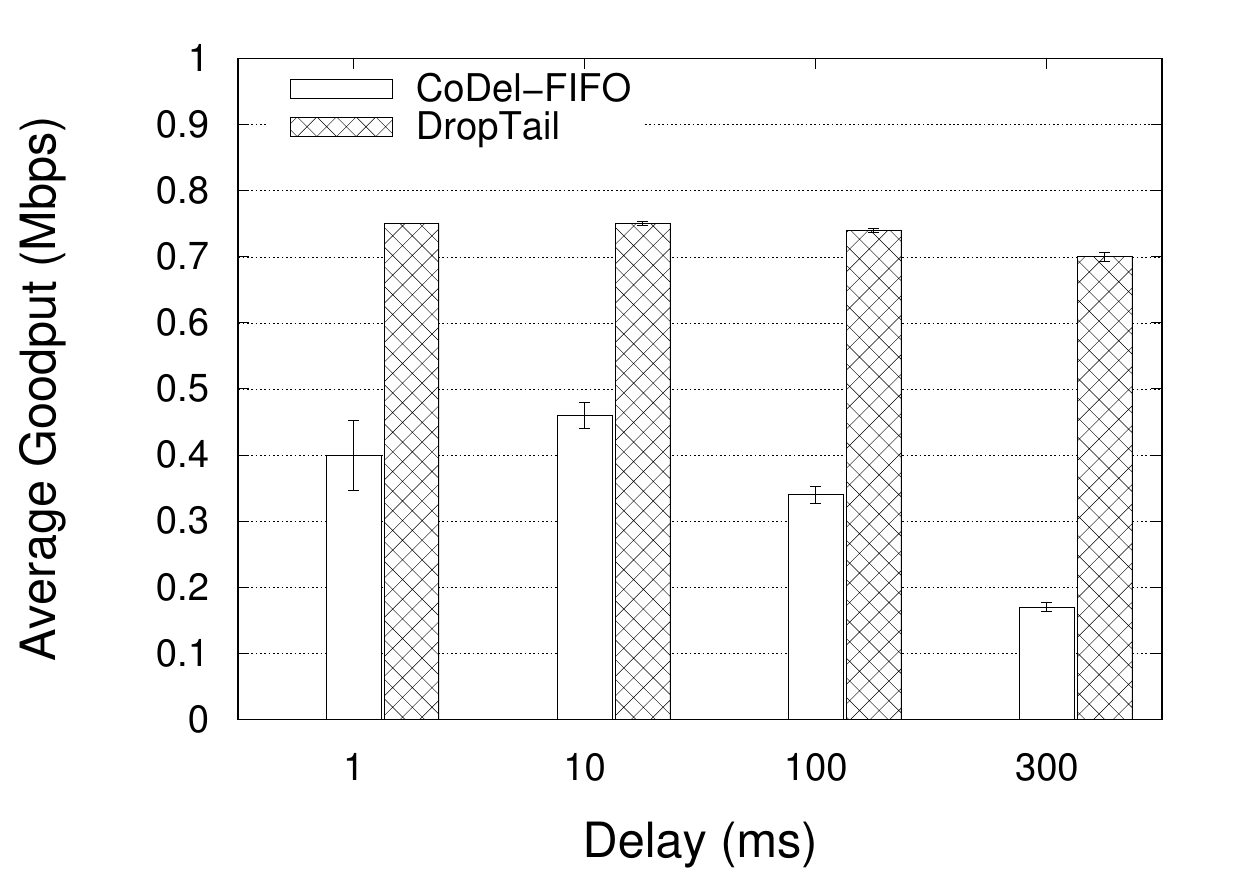}
        \caption{Goodput}
        \label{fig:codelgoodput}
    \end{subfigure}
    \caption{CoDel Results }
    \label{fig:queuestate}
    \vspace{-0.2cm}
\end{figure}

DropTail queue absorbs the traffic and this entails a large RTT without packet loss, characterizing the bufferbloat phenomenon. CoDel mitigates this problem. Fig.~\ref{fig:queuedrops} shows the average packet drops. CoDel drops increase as the LTE path delay increases. DropTail packet losses only occur when the queue is full. In this case, as delay increases, the queue size and packet losses reduce.
For all evaluated congestion control algorithms, simulation results show CoDel mechanism with higher number of drops.

CoDel reduces the RTT compared to DropTail. However Fig.~\ref{fig:codelgoodput} shows that CoDel has a large impact on the multipath transmission goodput. When both paths have the same delay ($1ms$), the goodput gets closer to $0.4$ Mbps, showing that CoDel queue drops degrade the transmission. Once congestion control algorithms perceive drops as a signal of packet loss or a congestion, it enters in the AIMD phase causing a goodput reduction. Congestion control algorithms are balanced with DropTail discipline. They use the full available bandwidth on bottleneck. CoDel presents a trade-off between goodput and latency (RTT). This was slightly perceived in~\cite{nichols:12}, but results show a more expressive difference with multipath congestion controls. These results motivated us to seek answering how to reduce this trade-off.

\subsection{CoDel-LIFO: Reducing Latency vs Goodput Trade-off}\label{subsec:fiforesults}

\begin{figure}[ht]
    \centering
        \begin{subfigure}[b]{0.24\textwidth}
        \centering
        \includegraphics[width=\textwidth]{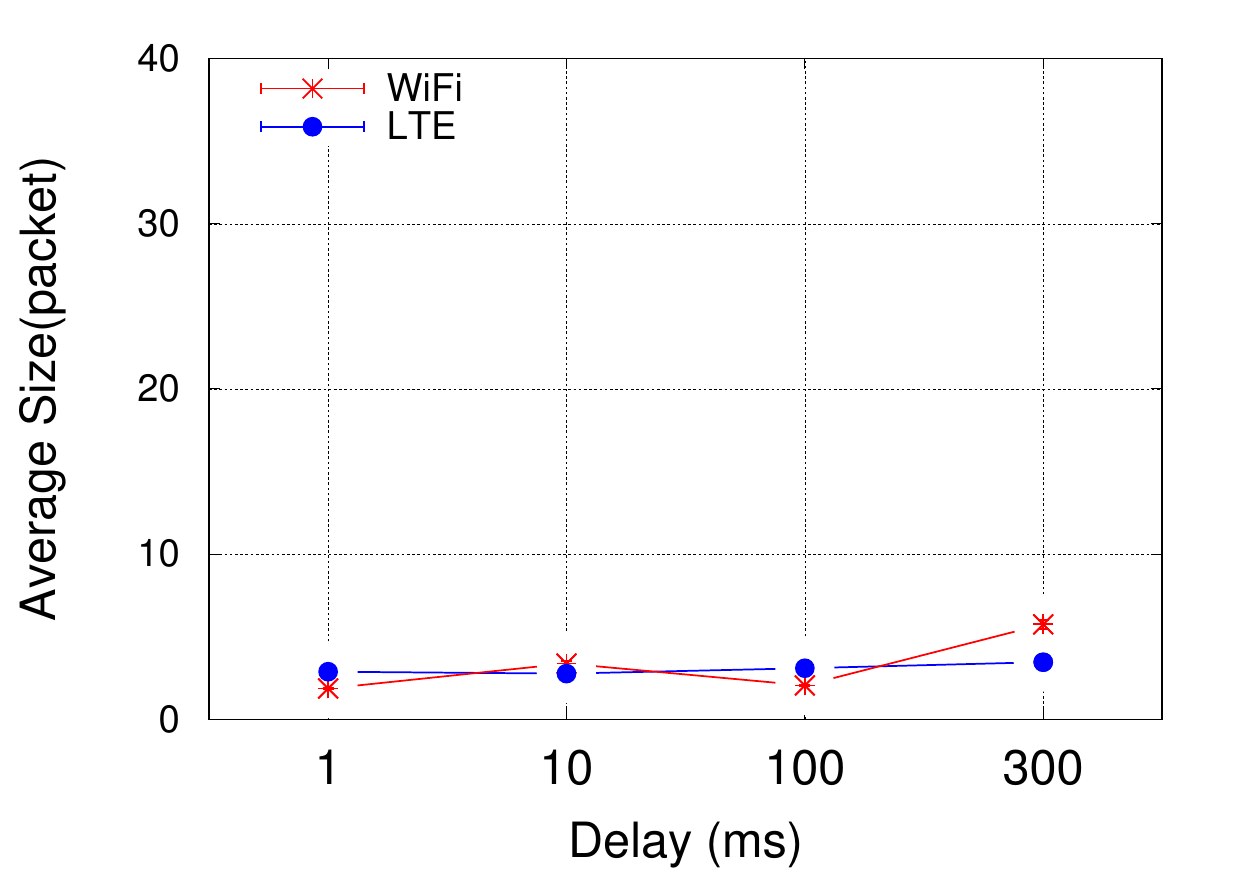}
        \caption{CWND}
        \label{fig:lifocwnd}
    \end{subfigure}
    \begin{subfigure}[b]{0.24\textwidth}
        \centering
        \includegraphics[width=\textwidth]{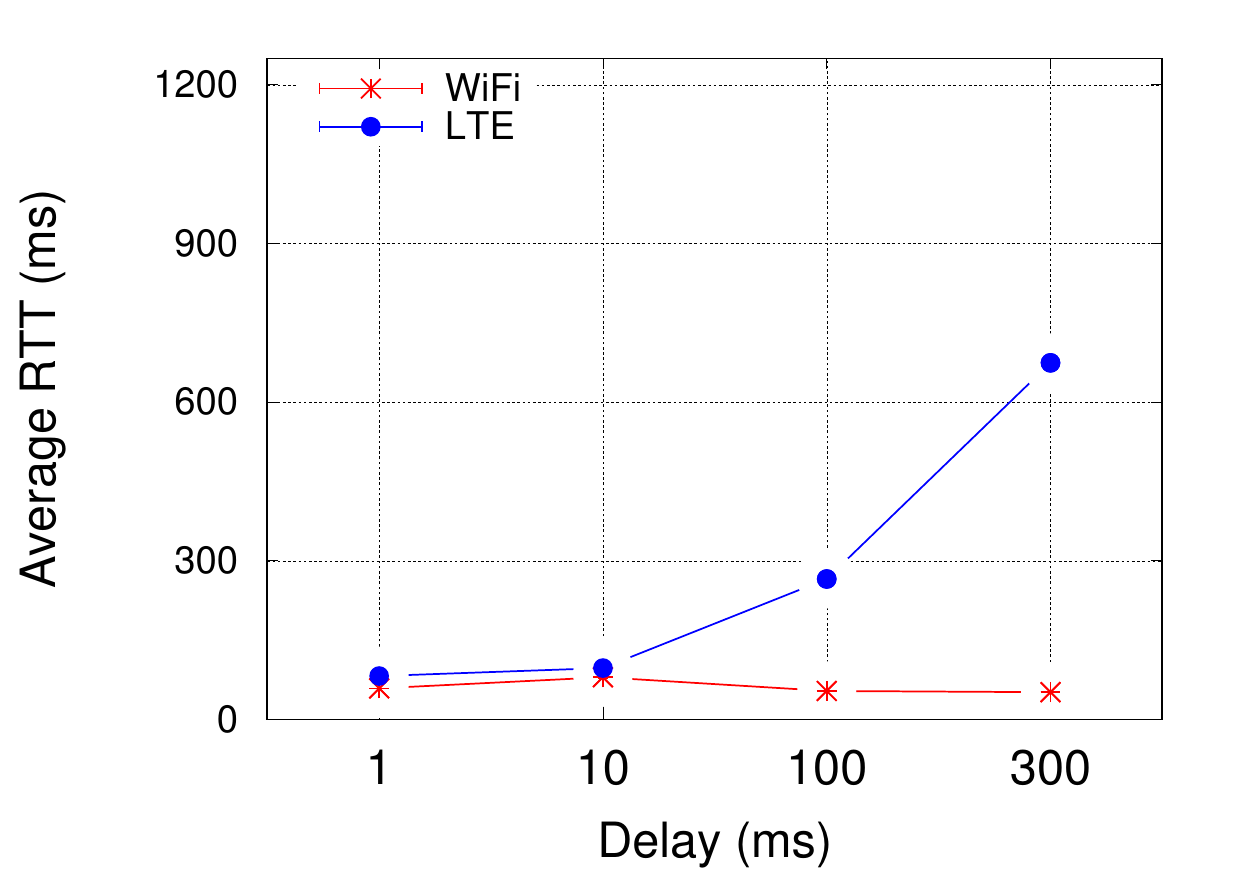}
        \caption{LIA average RTT}
        \label{fig:lifoliartt}
    \end{subfigure}
    
    \begin{subfigure}[b]{0.24\textwidth}
        \centering
        \includegraphics[width=\textwidth]{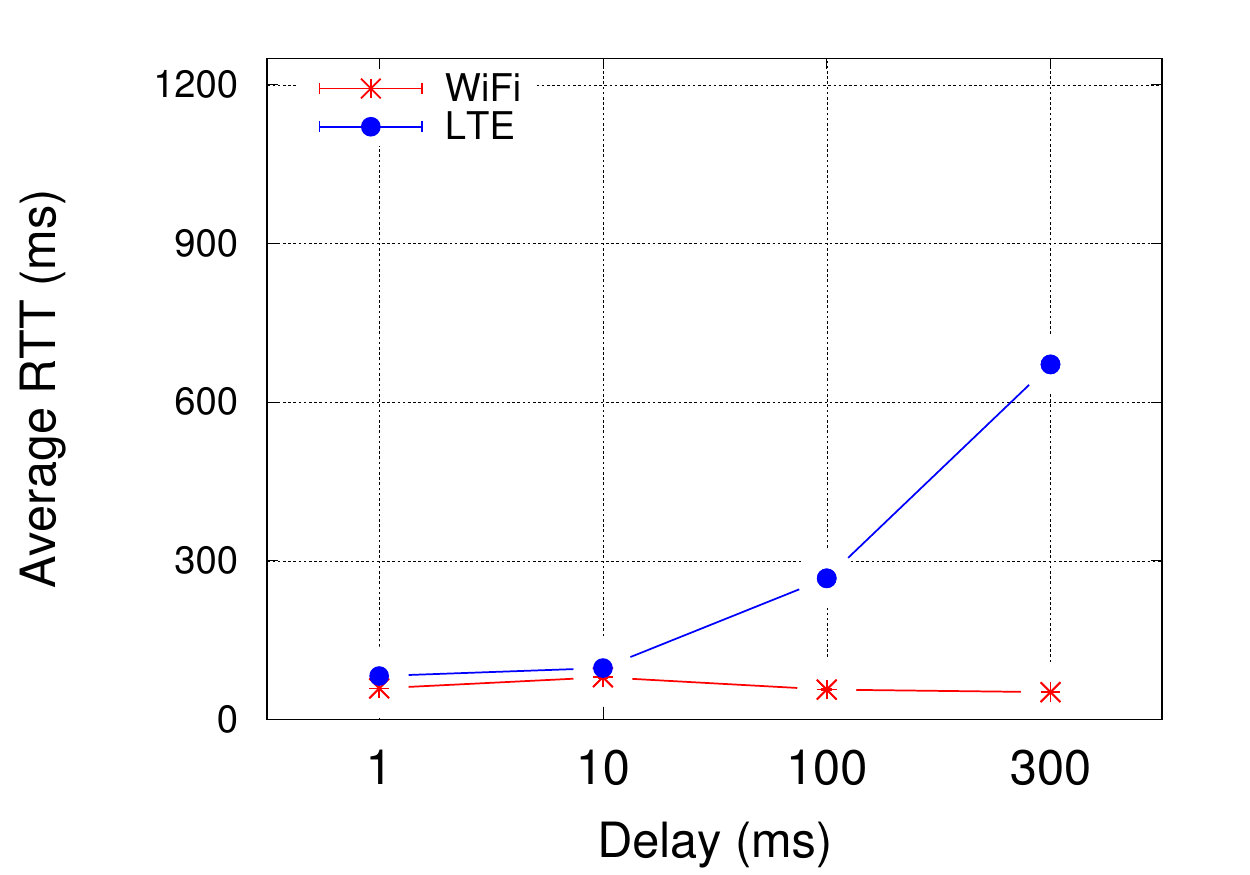}
        \caption{RTT Comp. average RTT}
        \label{fig:lifocprtt}
    \end{subfigure}
    \begin{subfigure}[b]{0.24\textwidth}
        \centering
        \includegraphics[width=\textwidth]{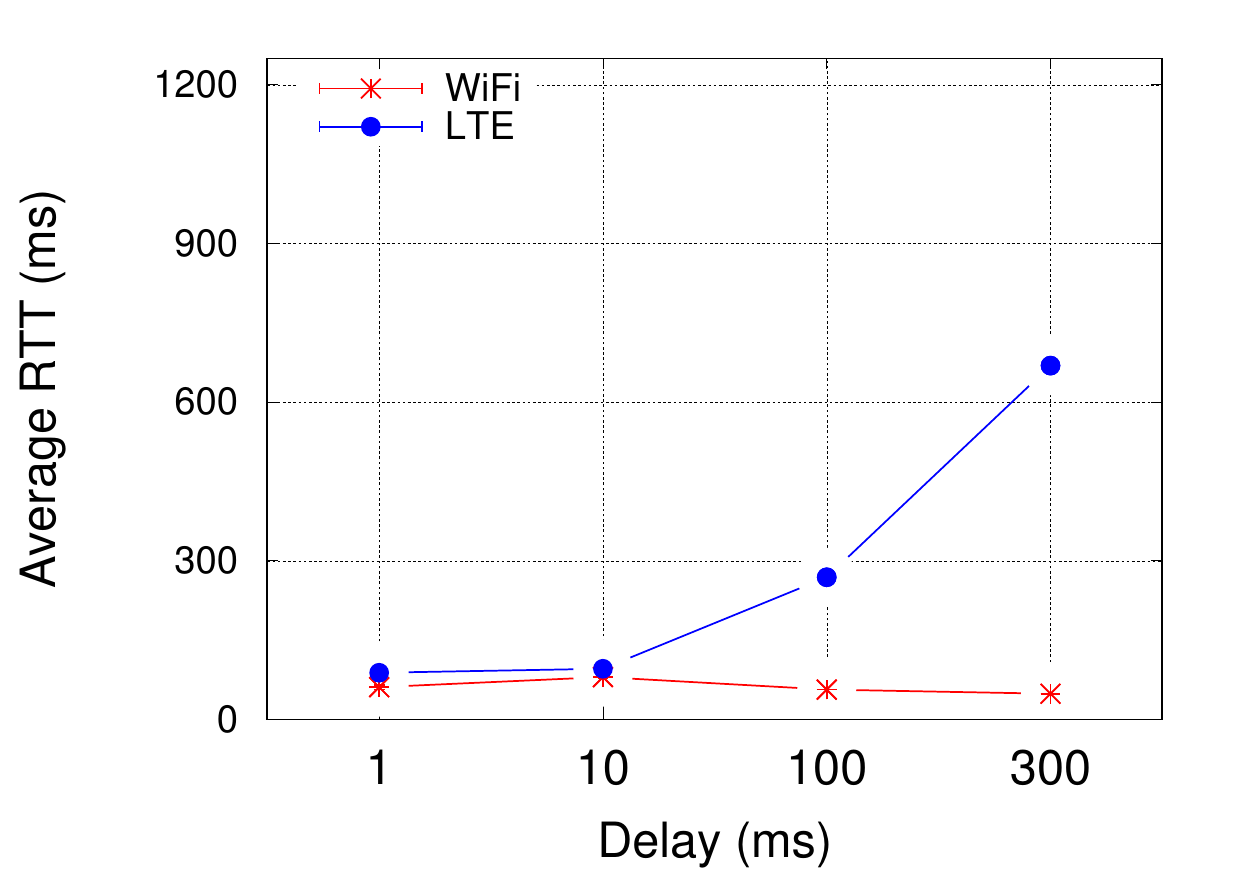}
        \caption{Uncoupled average RTT}
        \label{fig:lifouncrtt}
    \end{subfigure}
     
    \caption{CoDel-LIFO CWND and RTT results}
    \label{fig:liforttcwnd}
    \vspace{-0.5cm}
\end{figure}

This subsection presents simulation results employing CoDel-LIFO. Previous subsection showed the CoDel drawbacks, i.e. the trade-off between latency and goodput. CoDel-LIFO try to minimize this trade-off reducing the queue drops and prioritizing newer packets. Fig.~\ref{fig:lifocwnd} presents the average CWND of LIA with CoDel-LIFO as queue discipline. LIA has difficulties to move the traffic to path with lower RTT because CoDel-LIFO drops impact similarly in both paths. However, RTT on paths remains near to the result found over CoDel, as showing in Fig.~\ref{fig:liforttcwnd}. RTT remains low as in CoDel for all congestion control algorithms evaluated.

\begin{figure}[!ht]
\centering
\includegraphics[width=2.0in]{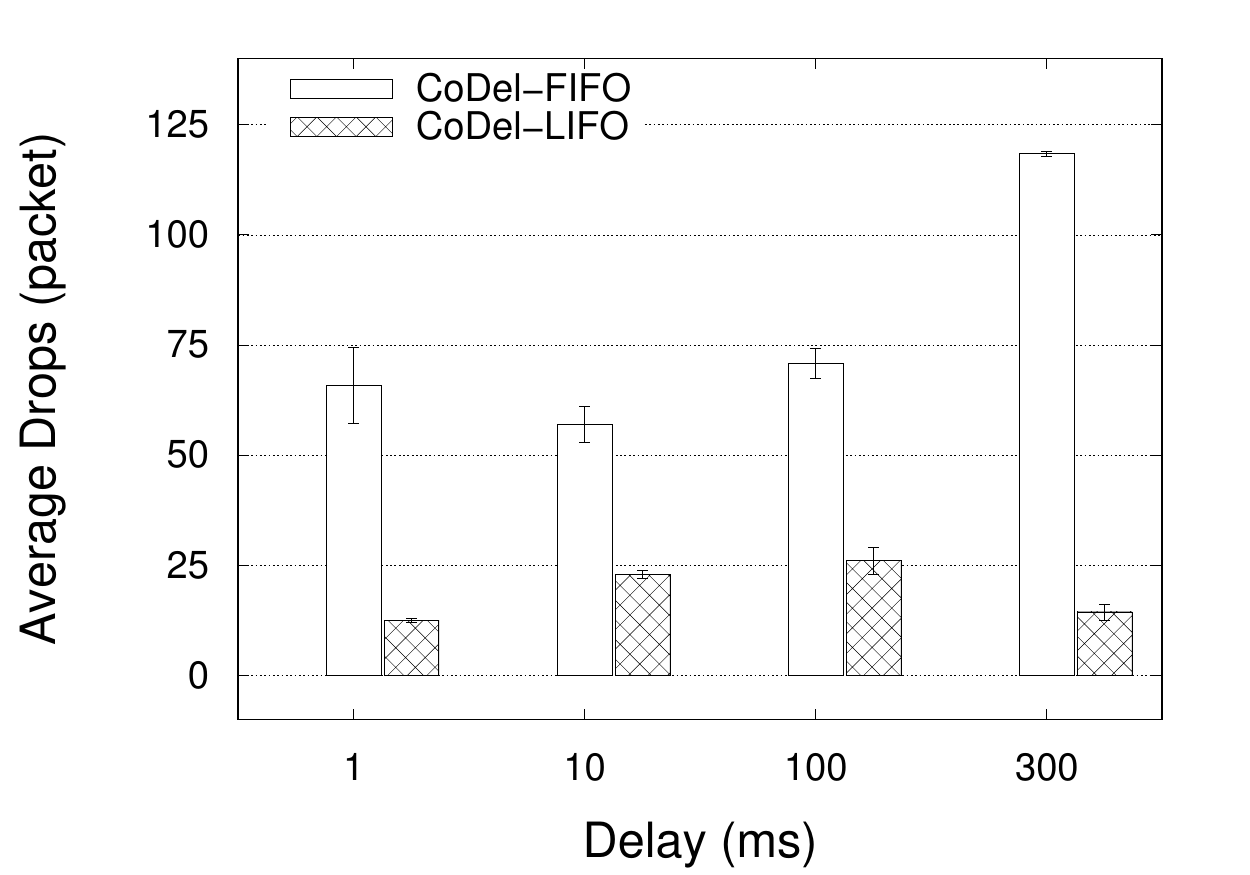}
\caption{CoDel-LIFO Drops}
\label{fig:lifodrops}
\vspace{-0.3cm}
\end{figure}

CoDel-LIFO drastically reduces the average queue drops. Drop reduction does not affect paths RTT, however it has more impact on average CWND size than CoDel. Fig.~\ref{fig:lifodrops} shows the CoDel-LIFO average queue drops comparing with CoDel. Average drop difference between both queue disciplines is greater. When the paths has the same RTT, CoDel-LIFO reduces in $6$ times the number of drops. With high variation in path delay, the reduction is up $10$ times.

\begin{figure}[ht]
    \centering
    \begin{subfigure}[b]{0.24\textwidth}
        \centering
        \includegraphics[width=\textwidth]{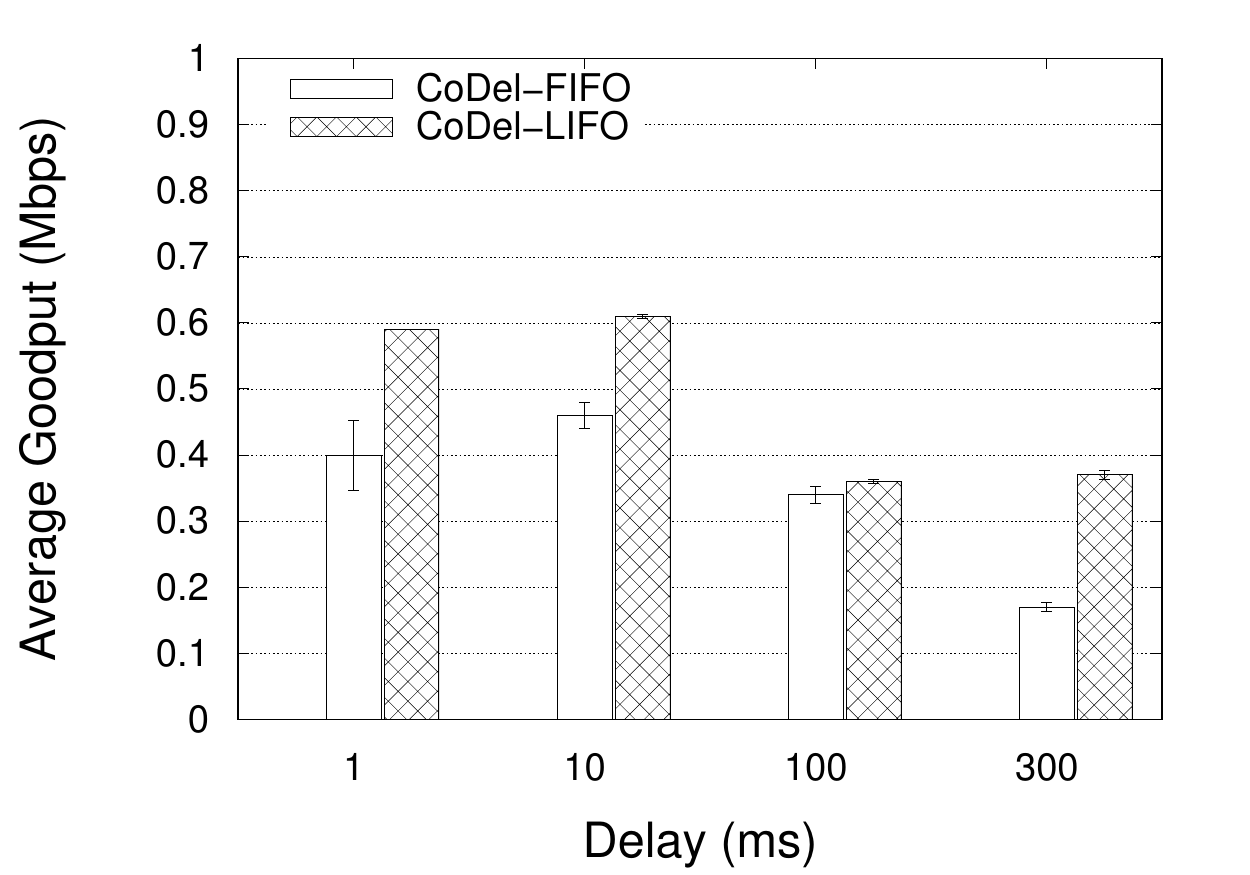}
        \caption{LIA Goodput}
        \label{fig:rttgoodput}
    \end{subfigure}
    \begin{subfigure}[b]{0.24\textwidth}
        \centering
        \includegraphics[width=\textwidth]{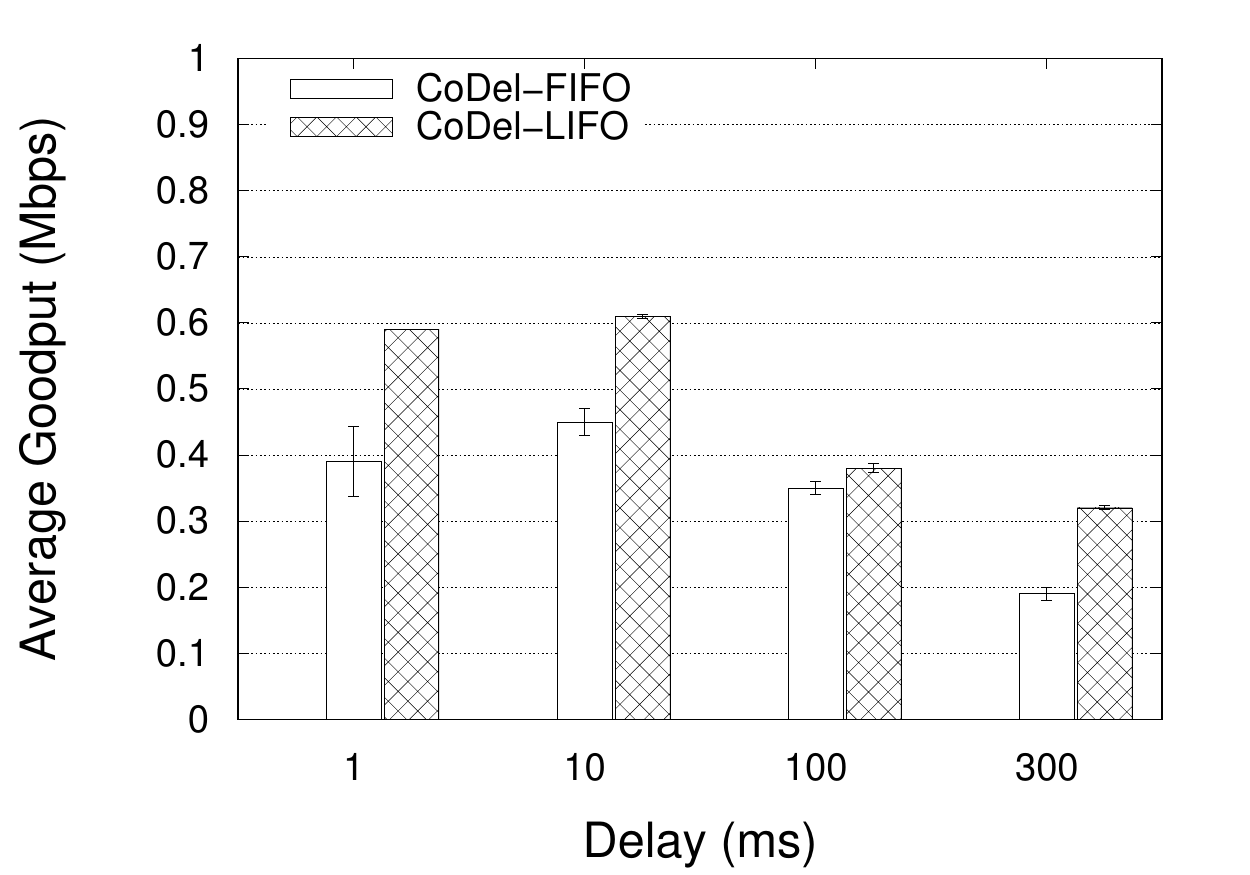}
        \caption{RTT Compensator Goodput}
        \label{fig:rttgoodput}
    \end{subfigure}

    \begin{subfigure}[b]{0.24\textwidth}
        \centering
        \includegraphics[width=\textwidth]{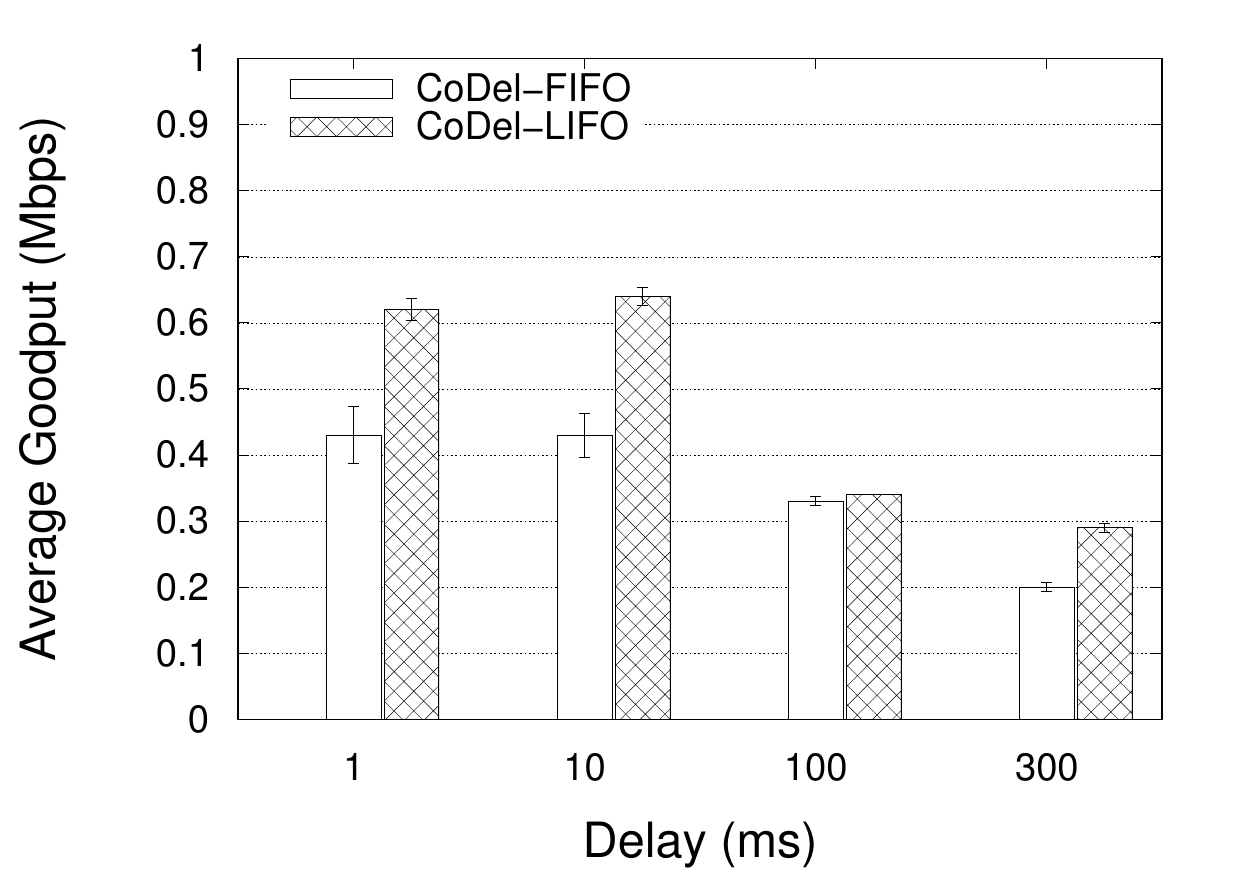}
        \caption{Uncoupled Goodput}
        \label{fig:uncgoodput}
    \end{subfigure}   
    
    \caption{CoDel-LIFO average goodput}
    \label{fig:lifogoodput}
    \vspace{-0.3cm}
\end{figure}

Drop reduction has less impact on congestion control algorithms. Fig.~\ref{fig:lifogoodput} shows the average goodput to congestion control algorithms and CoDel-LIFO as queue manage. Average goodput is at least $20\%$ higher than in CoDel. Increasing delay in path $A$, goodput degrades with both queue disciplines. This occurs because congestion control algorithms try prioritize the fast path reducing the send rate on slower path. However, even with drops impacting similarly in both paths, CoDel-LIFO goodput remains up to CoDel-FIFO results. This happens due to two things, the queue drop reduction and LIFO prioritization (i.e. stack). Queue drops fall due to $\theta$ limitation. RTT remains low because packets with shorter sojourn time are forwarded. Results with FQ-CoDel-LIFO and FQ-CoDel-FIFO  presents very close to what is observed with CoDel-FIFO and CoDel-LIFO. This is due to FQ-CoDel deployment uses independent queues per subflow. However, both subflows share the same queue. In this case, independent queues do not benefit from the multipath transmission. Hence, results are not shown.

Packet prioritization on stack structure reduces the mean sojourn time, and $\theta$ parameter contributes to queue length reduction. Fig.~\ref{fig:queuestate} shows the average queue length and average packet sojourn time with congestion control algorithms. LIA and RTT Compensator have similar impact on queue metrics. Uncoupled presents differences related to the others. This occurs because each subflows has an independent congestion window. It does not balances the traffic between the paths in order to prevent the RTT variations.

\begin{figure}[ht]
    \centering
    \begin{subfigure}[b]{0.24\textwidth}
        \centering
        \includegraphics[width=\textwidth]{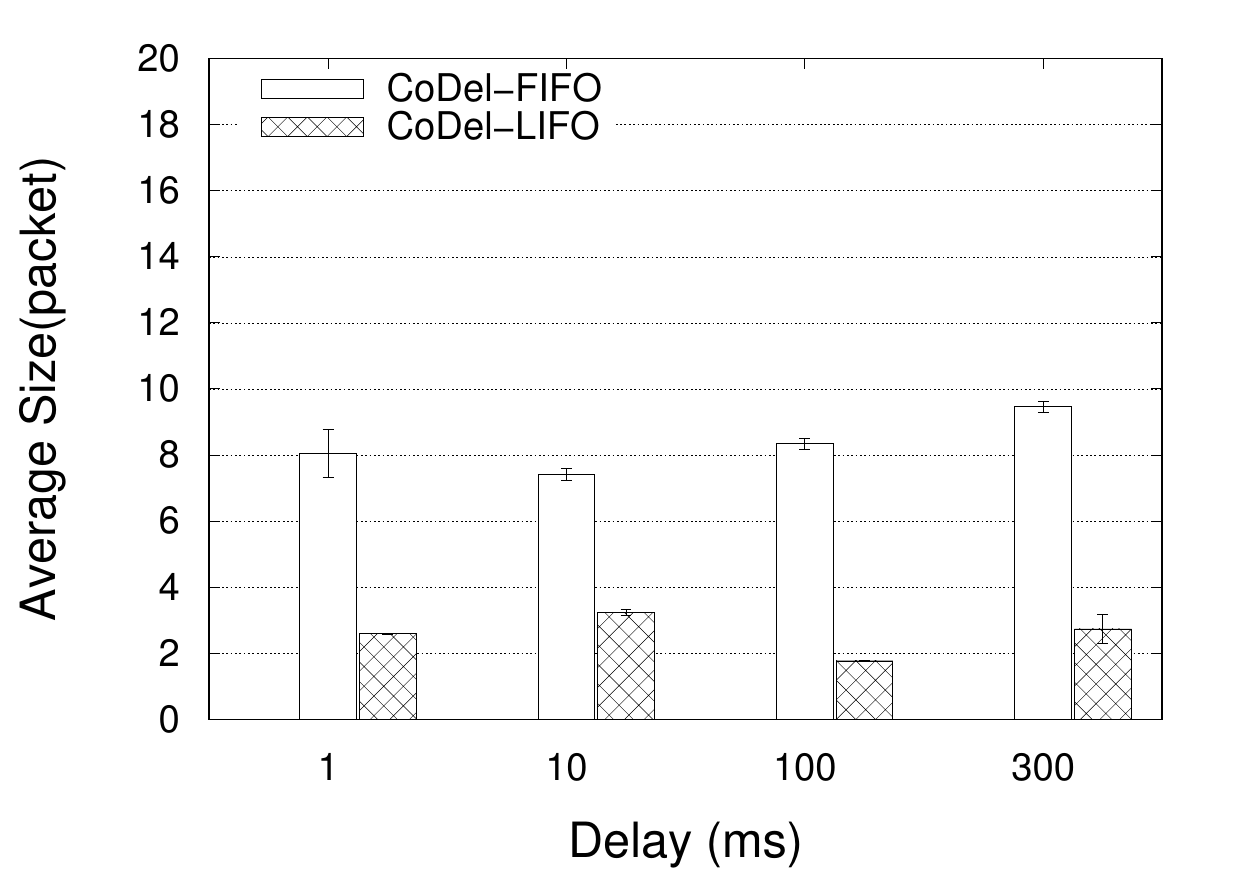}
        \caption{LIA Queue Length}
        \label{fig:lia-lght}
    \end{subfigure}
    \begin{subfigure}[b]{0.24\textwidth}
        \centering
        \includegraphics[width=\textwidth]{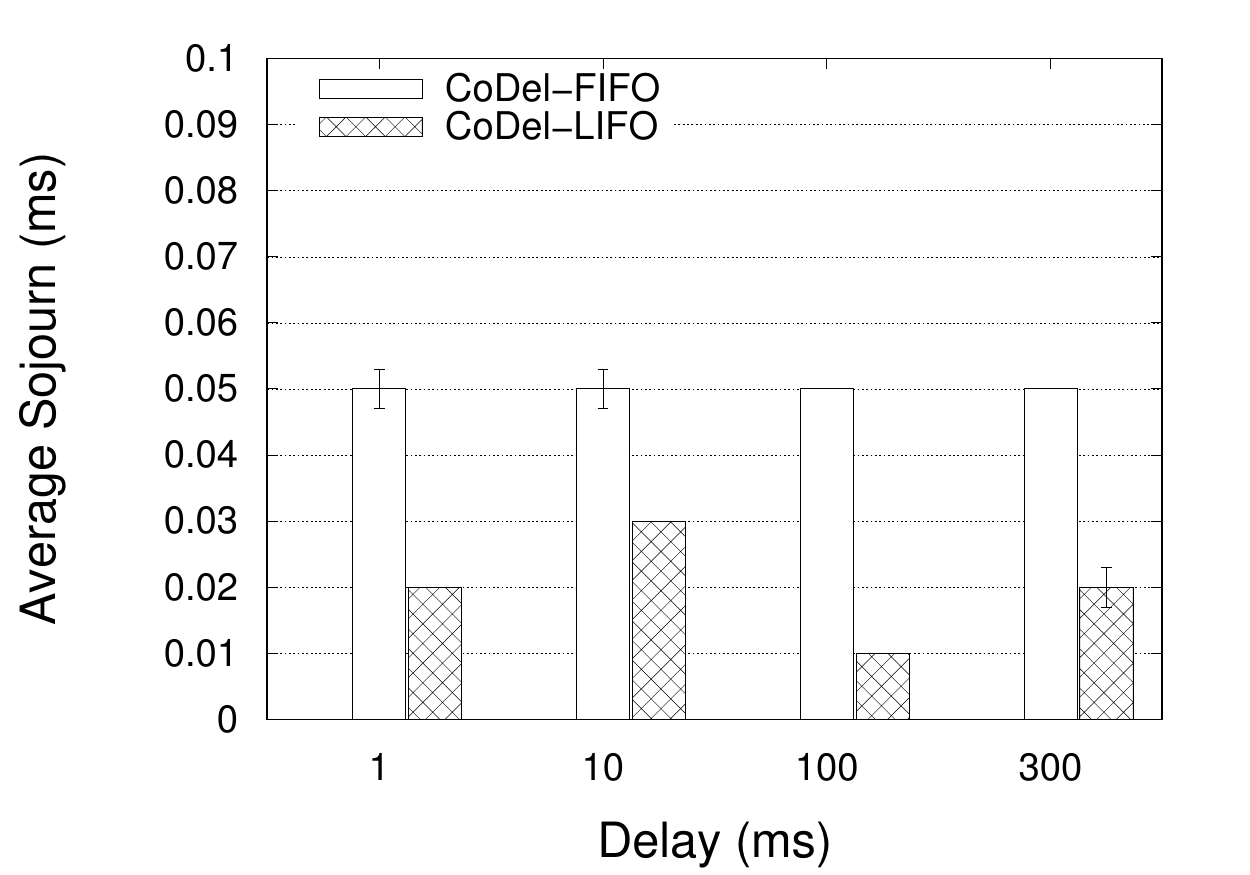}
        \caption{LIA Sojourn Time}
        \label{fig:lia-sjn}
    \end{subfigure}
    
    \begin{subfigure}[b]{0.24\textwidth}
        \centering
        \includegraphics[width=\textwidth]{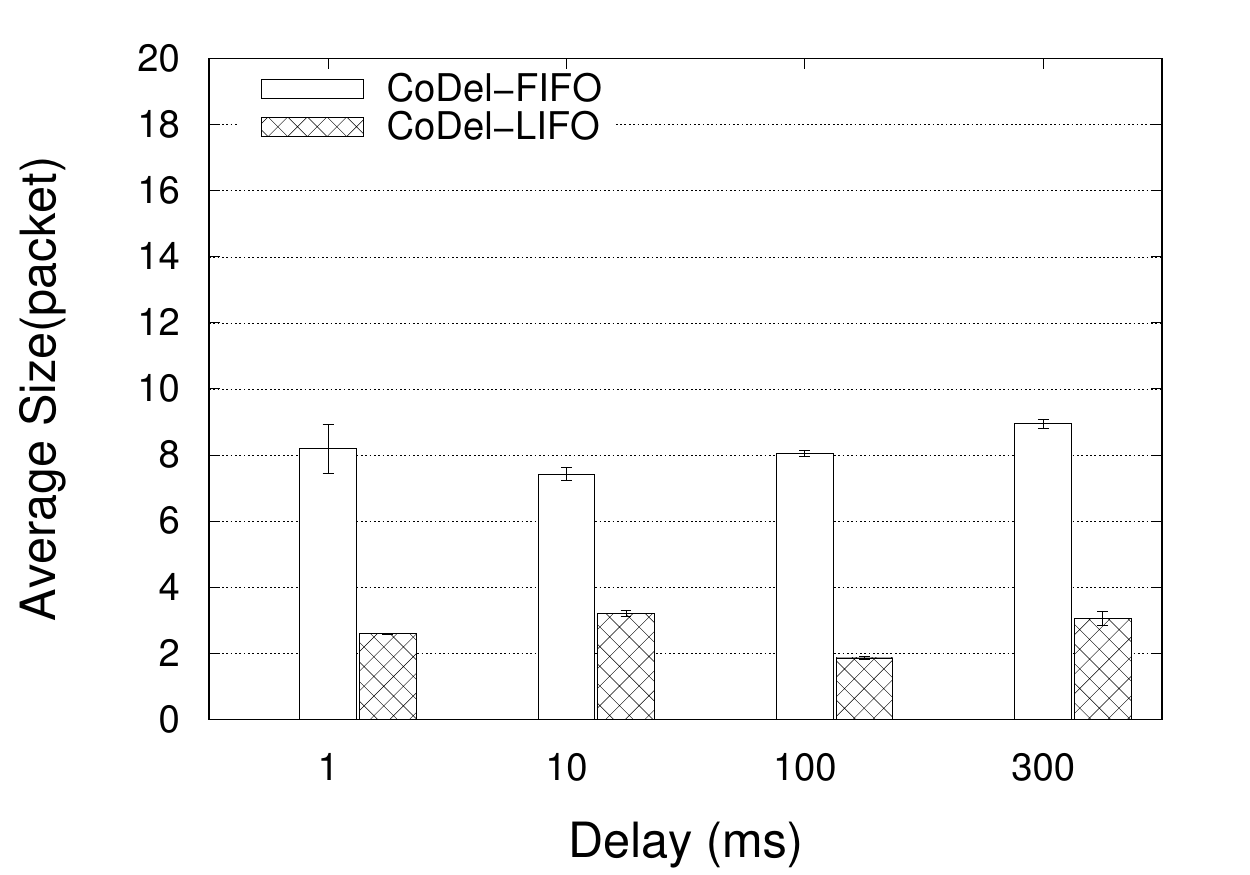}
        \caption{RTT Comp. Queue Length}
        \label{fig:rttc-lght}
    \end{subfigure}
    \begin{subfigure}[b]{0.24\textwidth}
        \centering
        \includegraphics[width=\textwidth]{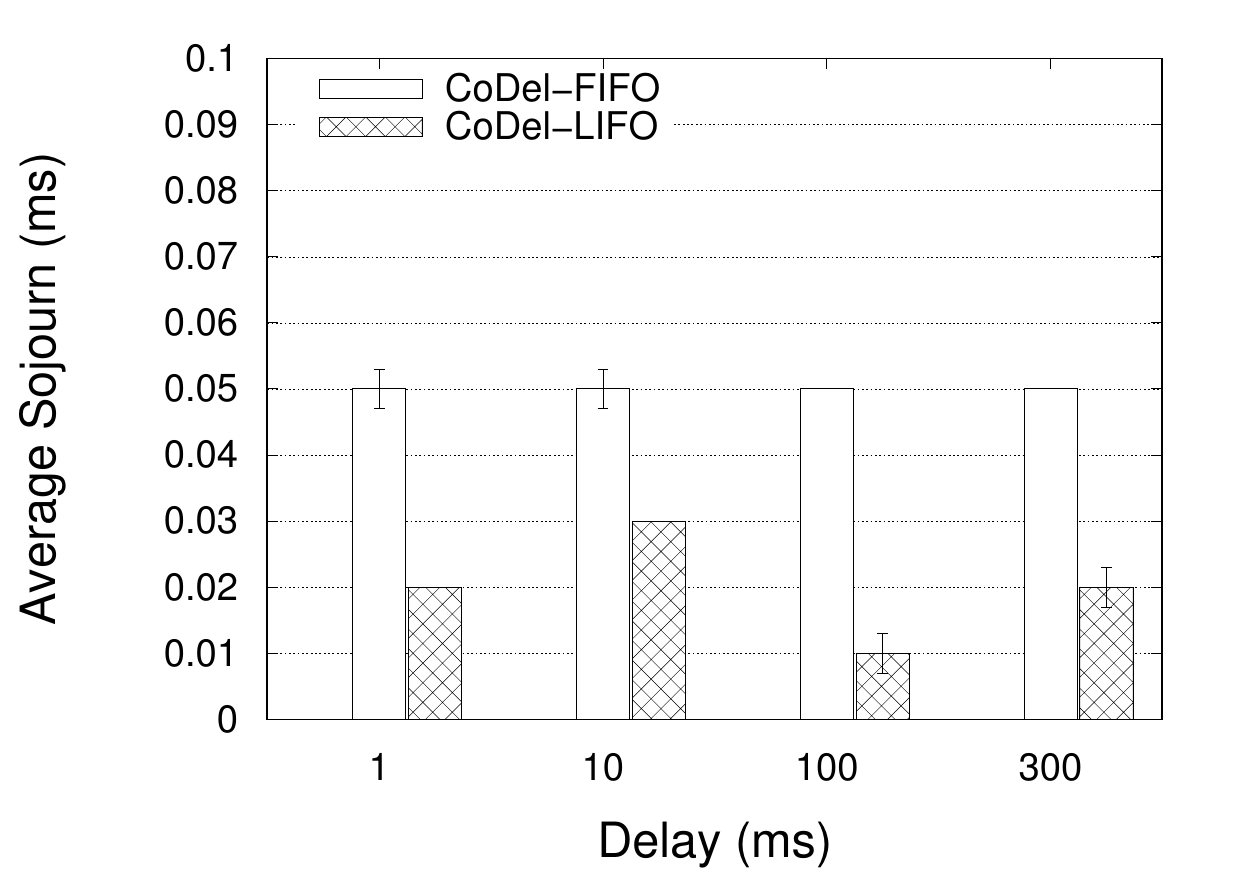}
        \caption{RTT Comp. Sojourn Time}
        \label{fig:rttc-sjn}
    \end{subfigure}

    \begin{subfigure}[b]{0.24\textwidth}
        \centering
        \includegraphics[width=\textwidth]{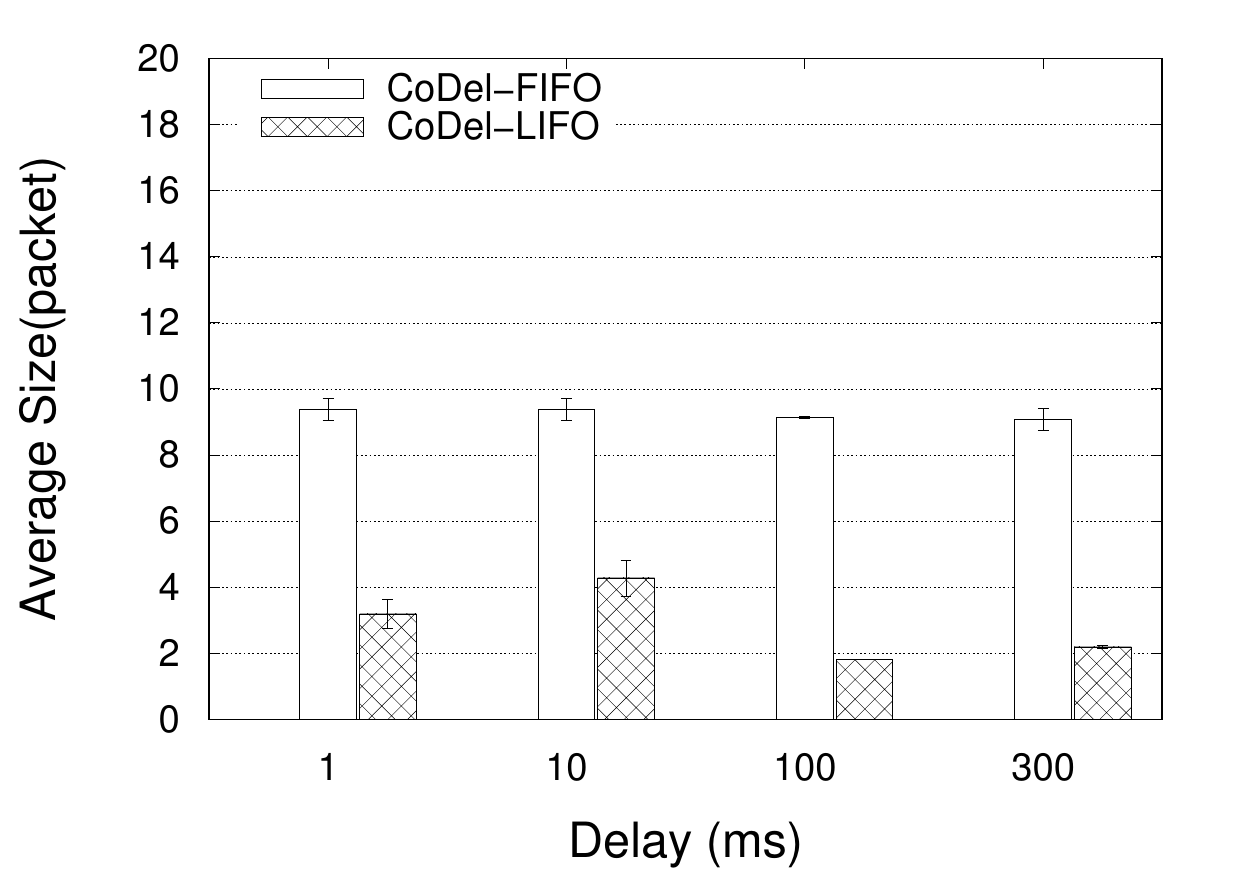}
        \caption{Uncoupled Queue Length}
        \label{fig:u-lght}
    \end{subfigure}
    \begin{subfigure}[b]{0.24\textwidth}
        \centering
        \includegraphics[width=\textwidth]{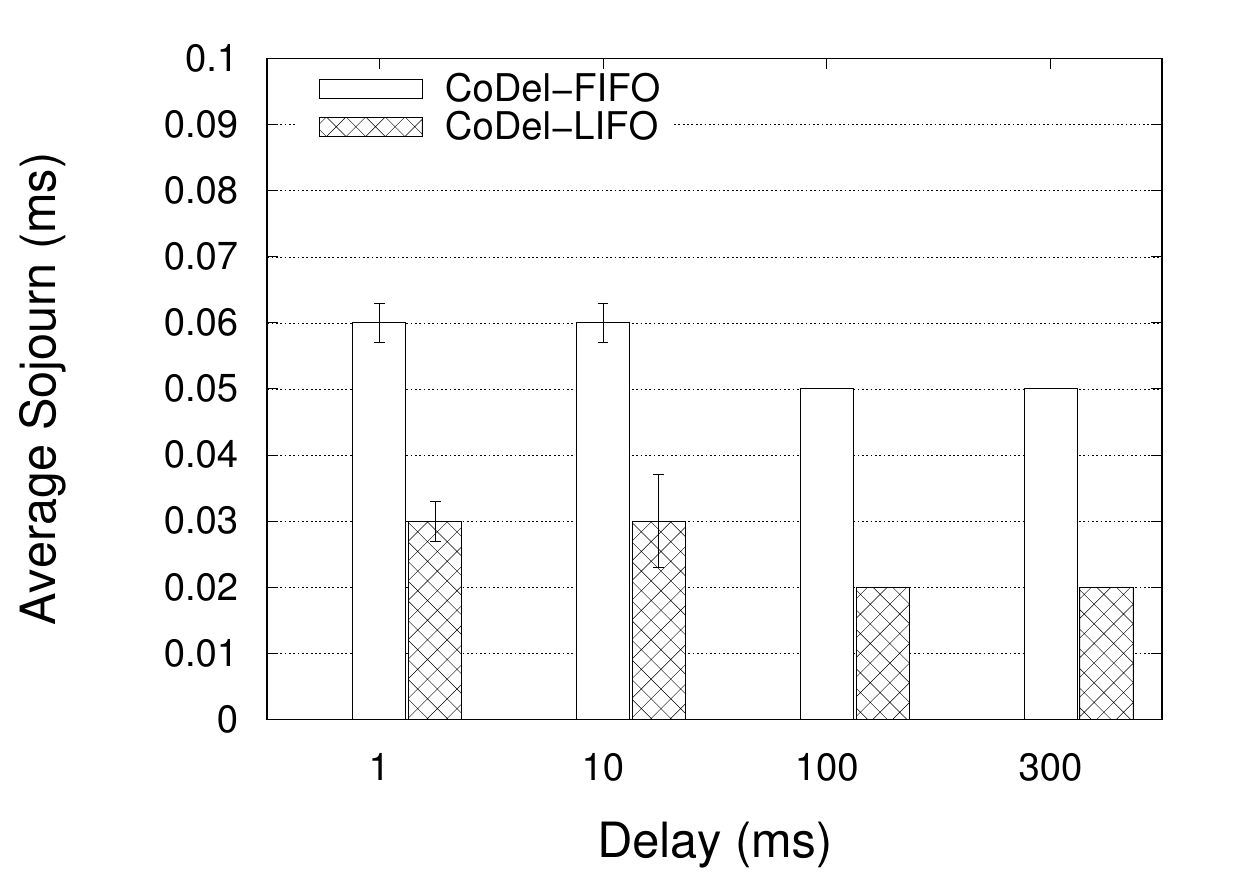}
        \caption{Uncoupled Sojourn Time}
        \label{fig:u-sjn}
    \end{subfigure}
    \caption{Average Queue Size and Sojourn Time}
    \label{fig:queuestate}
    \vspace{-0.5cm}
\end{figure}

\section{Related Work}\label{sec:related}

To the best of our knowledge, our work is the first to evaluate AQM impact on Multipath TCP congestion control. Solutions that addressed the bufferbloat problem are classified in end-node solutions and network solutions. End-node solutions are related to congestion control algorithms and network solutions to AQM disciplines. This work fills a gap showing that a hybrid (end-node and network) solution provides better results. We then open a new class of solutions, the hybrid ones.

Delay variation caused by network conditions has a great impact in concurrent multipath transfer. This is one of the main issues addressed by related works. Issues as HOL blocking and out-of-order arrive packets degraded whole transmission because added expressive delay to in-order delivery of packets to up-layer. These problems became worse by the presence of bufferbloat. Its effects have attracted high attention and motivated researchers to investigate their impact over multipath connections. Authors in~\cite{lee:15,ferlin14} investigated the bufferbloat effects on Multipath TCP in heterogeneous wireless networks. Lee and Lee~\cite{lee:15} proposed a dynamic adjustment of the receiver window based on the RTT value to overcome the effects. The work considered a scenario that depended on a range of RTT observations to work as planned. However, RTT variations in heterogeneous networks made difficult to deploy the proposal in practical scenarios.

Also, Ferlin~\emph{et al.} proposed an algorithm to mitigate the bufferbloat effects limiting the amount of data sent by congested path~\cite{ferlin14}. Results showed that the goodput was improved, but this could be ineffective when both paths were congested or shared a bottleneck. Chen and Towsley~\cite{chen14} showed that the fastest path went through several periods of idleness due to packets arriving out of order, mainly when the slowest path experienced bufferbloat. This issue made MPTCP to restart the window in slow start mode causing performance degradation. The authors suggested to disable the mechanism in order to make the window to continue to grow after idle periods. This amendment improved the performance, but MPTCP subflows became more aggressive over TCP flows.

The loss rate prevented the growth of the congestion window and thus the transmission could not reach the full capacity of the transmission channel. Gomez~\emph{et al.} performed simulations comparing different MPTCP CC algorithms under different link loss rate (Frame Error Rate)~\cite{gomez14}. Results showed the performance degradation of MPTCP CC under increasing packet loss. However, the work only considered losses in access network and did not extend the analysis to losses caused by events in the network core, such AQM disciplines.

Related to AQMs, authors in~\cite{kulatunga:15} evaluated CoDel and FQ-CoDel over networks with limited transmission capacity and high RTTs. In order to reduce latency, they optimized the CoDel parameters for a specific scenario. However, this optimization could be impractical in dynamic scenarios where the parameters needed to be adjusted frequently. 

Havey and Almeroth~\cite{havey:15} showed that traditional AQM, like CoDel and PIE, did not protect Layer 2 (MAC) bufferbloat. Authors showed the bufferbloat existence in queues below the IP layer. AQM algorithms could not detect this issue. To solve this an Active Sense Queue Management (ASQM) to solve this problem. ASQM run at the IP layer, but, sensed entire access link to determine the queuing delay across the Link layer and sent a mark/drop signal to TCP sender slow down. Its controls bufferbloat on par with IP layers AQM (CoDel and PIE) to provide a better bufferbloat protection. Our approach was suitable to work with ASQM as well as CoDel. 

Gong et al.~\cite{gong:14} evaluated the fairness of the capacity shared between best-effort TCP and Low Priority Congestion Control (LPCC) flown on a bottleneck governed by AQM. Results showed that AQMs (CoDel, RED, etc) redefined the priority level between best-effort TCP and LPCC. On a bottleneck TCP share reduced dramatically, becoming close to the LPCC share. Our approach evaluated the subflows fairness impact with separate queues (FQ-CoDel) approach. However, left out subflows priority.

The works mentioned above assess the bufferbloat in HetNets with Multipath TCP, but none of them had analyzed the congestion control algorithms under AQM disciplines and bufferbloat jointly in Heterogeneous Wireless Networks.

\section{Conclusion}\label{sec:conclusion}

This work analyzed the impact of AQM disciplines in Multipath TCP congestion control in heterogeneous wireless networks scenarios. Simulation results showed that default queue management (DropTail) caused long queue delays. In our investigations, we showed that CoDel and FQ-CoDel mitigated this issue controlling the queue size and reducing the delay. However, we showed that CoDel caused many queue drops and its impact on multipath congestion control algorithms, causing a goodput degradation of multipath transmission. The proposed CoDel-LIFO reduced the packet loss and this improved the transmission increasing the goodput without compromise the latency. The improvement on multipath transmission was achieved by changing queue to stack and forwarding packets with large sojourn time. CoDel-LIFO kept the queue length and the sojourn time smaller than CoDel-FIFO and this reduced the mean sojourn time. The $\theta$ parameter controlled packet forwarding with large sojourn time based on the difference between early and current sojourn time. However, our investigations were limited to one multipath connection. As a future work, we intend to evaluate delay-based multipath congestion control algorithms, buffer reorder overhead and spurious re-transmission impact.

\section*{Acknowledgment}
The authors would like to thank the Wireless and Advanced Networks (NR2) team and the CAPES for supporting this research.

\bibliographystyle{IEEEtran}
\bibliography{IEEEACM-ToN}
\end{document}